\documentclass[review, 12pt]{elsarticle}
\usepackage{lineno,hyperref,amsmath,amsthm}
\usepackage[noend]{algpseudocode}
\usepackage{algorithmicx,algorithm}
\usepackage{booktabs}
\usepackage{graphics}
\usepackage{subfig}
\modulolinenumbers[1]

\journal{Journal} %

\begin{document}

\begin{frontmatter}

\title{A new numerical scheme for simulating non-gaussian and non-stationary stochastic processes}

\author[a,b]{Zhibao Zheng\corref{CorrespondingAuthor}}
\cortext[CorrespondingAuthor]{Corresponding author.}
\ead{zhibaozheng@hit.edu.cn}

\author[a,b]{Hongzhe Dai}

\author[a,b]{Yuyin Wang}

\author[a,b]{Wei Wang}

\address[a]{Key Lab of Structures Dynamic Behavior and Control, Harbin Institute of Technology, Ministry of Education, Harbin 150090, China}

\address[b]{School of Civil Engineering, Harbin Institute of Technology, Harbin 150090, China}

\begin{abstract}
This paper presents a new numerical scheme for simulating stochastic processes specified by their marginal distribution functions and covariance functions. Stochastic samples are firstly generated to automatically satisfy target marginal distribution functions. An iterative algorithm is proposed to match the simulated covariance function of stochastic samples to the target covariance function, and only a few times iterations can converge to a required accuracy. Several explicit representations, based on Karhunen-Lo\`{e}ve expansion and Polynomial Chaos expansion, are further developed to represent the obtained stochastic samples in series forms. Proposed methods can be applied to non-gaussian and non-stationary stochastic processes, and three examples illustrate their accuracies and efficiencies.
\end{abstract}


\end{frontmatter}


\section{Introduction}

With widely developments of uncertainty quantification theories and methods, stochastic problems involving uncertainties commonly arise in various fields of engineering, such as computational mechanics \cite{ghanem2003stochastic}, financial analysis \cite{tankov2003financial} and biomedical science \cite{aboy2005adaptive}. A large number of these problems involves uncertain quantities which should be modeled as random processes or fields. On the one hand, assumptions regarding probabilistic distributions are made due to the incomplete experimental data \cite{stefanou2009stochastic}. On the other hand, stochastic simulations are provided for sufficient observation data. Thus, applications of stochastic process and field theories to engineering problems have gained considerable interests. In general, stochastic processes are assumed to be gaussian because of simplicity and the Central Limit Theorem. Since the Gaussian stochastic processes can be completely described by their second-order statistics, \cite{papoulis2002probability}, methods for simulating gaussian stochastic processes \cite{grigoriu2006evaluation, rasmussen2003gaussian, phoon2004comparison} have been quite well established. However, the gaussian assumption does not work due to the fact that some physical phenomena are obviously non-gaussian in some cases \cite{deodatis2001simulation}. The difficulty for simulating non-gaussian stochastic processes is that all the joint distribution functions are needed to completely characterize the non-gaussian properties. The problem is even further complicated when the process is also non-stationary since the marginal distributions depend on time or space. With these motivations, the efficient simulation of non-gaussian stochastic processes are urgent because of practical and theoretical importance.

Spectral representation is the first widely developed method for the simulation of non-gaussian stochastic processes \cite{yamazaki1988digital}. This method is implemented in frequency domain and is initially developed for gaussian stochastic processes \cite{shinozuka1991simulation}. It has been extended to non-gaussian stochastic processes by combining the spectral representation method with non-linear transformations \cite{yamazaki1988digital}, i.e., tranforming gaussian stochastic samples generated by the spectral representation method into the non-gaussian stochastic process and matching the target power spectral density function and non-gaussian marginal distribution function. Extensive studies based on this method can be found in \cite{Elishakoff1994Conditional, Popescu1998Simulation, grigoriu1998simulation, liu2016random}. Different from the spectral representation method, Karhunen-Lo\`{e}ve (KL) expansion \cite{ghanem2003stochastic, huang2001convergence} is implemented in time or space domains, which is usually used in the simulation of stationary and non-stationary gaussian processes \cite{sudret2000stochastic, poirion2013non, kim2015modeling, liu2017dimension}.
Iterative algorithms for updating the non-gaussian expanded random variables are proposed in \cite{phoon2002simulation, phoon2005simulation} for the simulation of non-gaussian stochastic processes.  The method can be applied to highly skewed non-gaussian marginal distribution functions. Hence, KL expansion provides a unified and powerful framework for the simulation of stochastic processes, which is potentially capable of providing a better fit to non-gaussian and non-translational data \cite{dai2019explicit}. Another important technique, Polynomial Chaos (PC) expansion, has also been developed for simulation of non-gaussian and non-stationary stochastic processes and fields in \cite{sakamoto2002polynomial, sakamoto2002simulation}. The method represents the target stochastic process and field as multidimensional Hermite polynomial chaos in a set of normalized gaussian random variables. The accuracy and efficiency of this method were further examined in \cite{puig2002non, field2004accuracy}.

In this paper, we present numerical schemes for simulating non-gaussian and non-stationary stochastic processes that have been specified by their covariance functions and non-gaussian marginal distribution functions. The basic idea is to firstly generate stochastic samples that satisfying target marginal distribution functions, and then match target covariance functions by developing delicate iterative algorithm. In this way, the simulation of both gaussian and non-gaussian stochastic processes  can be implemented in an unified framework since marginal distribution functions are automatically satisfied by generated samples, and the accuracy and efficiency of the simulation are only dependent on matching the target covariance functions.
Another advantage is that there are no differences in the application of the proposeed iterative algorithm to stationary and non-stationary stochastic processes. Thus, the proposed method can be considered as a unified numerical scheme for simulating samples of stochastic processes. Further, it's usually not convenient to apply stochastic samples in practical problems. In this paper, we exploit KL expansion for expanding the obtained stochastic samples since KL expansion is optimal among series expansion methods in the global mean square error with respect to the number of random variables in the representation. Thus, the proposed strategy is capable of representing stochastic processes with sufficient accuracy with as few random variables as possible. In order to meet the requirements of different practical problems, we also exploit PC expansion and KL-PC expansion (combination of KL expansion and PC expansion) to represent the obtained stochastic samples, whose methodology are similar to KL expansion but based on different expansions. The accuracies and efficiencies are
demonstrated by several numerical examples. Proposed methods can be readily generalized to multi-dimensional random fields  \cite{sakamoto2002simulation, lagaros2005enhanced, zheng2017simulation, betz2014numerical, christou2016optimal}, but it's beyond the scope of this article and will be studied in subsequent papers.

The paper is organized as follows: a new algorithm for simulating stochastic samples is presented in Section \ref{Sec2}, Section \ref{Sec3} develops several numerical algorithms for representing the obtained stochastic samples and three illustrative examples are finally given in Section \ref{Sec4} to demonstrate the proposed algorithms.

\section{Simulation of stochastic samples} \label{Sec2}

Consider a stochastic process $\omega \left( {x,\theta } \right)$, $x \in D$ specified by its covariance function $C\left( {{x_1},{x_2}} \right)$ and marginal distribution function $F\left( {y;x} \right)$. 
In order to obtain stochastic samples that satisfying the target covariance function and marginal distribution function, we discretize spatial domain as $x = \left\{ {{x_1}, \cdots ,{x_n}} \right\}$ and generate random variables samples $\left\{ {\left\{ {{\eta _i}\left( {{\theta _k}} \right)} \right\}_{k = 1}^N} \right\}_{i = 1}^n$ according to the target marginal distribution function $F\left( {y;x} \right)$, where $N$ is the number of random variables and ${\eta_i}\left( \theta  \right) = \omega \left( {{x_i},\theta } \right), ~i = 1, \cdots, n$.
Note that random variables $\left\{ {{\eta _i}\left( \theta  \right)} \right\}_{i = 1}^n$ automatically satisfy the target marginal distribution function, i.e., ${\eta _i}\left( \theta  \right) \sim F\left( {y;{x_i}} \right)$.
However, the generated samples of random variables $\left\{ {{\eta _i}\left( \theta  \right)} \right\}_{i = 1}^n$ don't match the target covariance functions $C\left( {{x_1},{x_2}} \right)$ since statistical correlations between samples $\left\{ {{\eta _i}\left( {{\theta _k}} \right)} \right\}_{k = 1}^N$ and $\left\{ {{\eta _j}\left( {{\theta _k}} \right)} \right\}_{k = 1}^N$ are no constraints.
Thus, delicate algorithm is required to be developed to match the target covariance function $C\left( {{x_1},{x_2}} \right)$.

Statistical correlations between random variable ${\eta _i}\left( \theta  \right)$ and ${\eta _j}\left( \theta  \right)$ are given as
\begin{equation}\label{Tij}
  {T_{ij}} = \frac{1}{{N - 1}}\sum\limits_{k = 1}^N {\left[ {{\eta _i}\left( {{\theta _k}} \right) - {{\bar \eta }_i}} \right]\left[ {{\eta _j}\left( {{\theta _k}} \right) - {{\bar \eta }_j}} \right]}
\end{equation}
where ${\bar \eta _i}$ and ${\bar \eta _j}$ are the mean of random variables ${\eta _i}\left( \theta  \right)$ and ${\eta _j}\left( \theta  \right)$, respectively. Expanding Eq.\eqref{Tij} yields,
\begin{align}\label{Tij_M}
{T_{ij}} &= \frac{1}{{N - 1}}\sum\limits_{k = 1}^N {\left( {{\eta _i}\left( {{\theta _k}} \right){\eta _j}\left( {{\theta _k}} \right) - {\eta _i}\left( {{\theta _k}} \right){{\bar \eta }_j} - {{\bar \eta }_i}{\eta _j}\left( {{\theta _k}} \right) + {{\bar \eta }_i}{{\bar \eta }_j}} \right)}\nonumber\\
&= \frac{1}{{N - 1}}\left[ {\sum\limits_{k = 1}^N {{\eta _i}\left( {{\theta _k}} \right){\eta _j}\left( {{\theta _k}} \right)}  - \left( {\sum\limits_{k = 1}^N {{\eta _i}\left( {{\theta _k}} \right)} } \right)\left( {\frac{1}{N}\sum\limits_{k = 1}^N {{\eta _j}\left( {{\theta _k}} \right)} } \right)} \right.\nonumber\\
&\left. { - \left( {\frac{1}{N}\sum\limits_{k = 1}^N {{\eta _i}\left( {{\theta _k}} \right)} } \right)\left( {\sum\limits_{k = 1}^N {{\eta _j}\left( {{\theta _k}} \right)} } \right) + N\left( {\frac{1}{N}\sum\limits_{k = 1}^N {{\eta _i}\left( {{\theta _k}} \right)} } \right)\left( {\frac{1}{N}\sum\limits_{k = 1}^N {{\eta _j}\left( {{\theta _k}} \right)} } \right)} \right]\nonumber\\
&= \frac{1}{{N - 1}}\left[ {\sum\limits_{k = 1}^N {{\eta _i}\left( {{\theta _k}} \right){\eta _j}\left( {{\theta _k}} \right)}  - \frac{1}{N}\left( {\sum\limits_{k = 1}^N {{\eta _i}\left( {{\theta _k}} \right)} } \right)\left( {\sum\limits_{k = 1}^N {{\eta _j}\left( {{\theta _k}} \right)} } \right)} \right]\nonumber\\
&= \frac{1}{{N - 1}}\sum\limits_{k = 1}^N {{\eta _i}\left( {{\theta _k}} \right){\eta _j}\left( {{\theta _k}} \right)}  - \frac{1}{{N\left( {N - 1} \right)}}\left( {\sum\limits_{k = 1}^N {{\eta _i}\left( {{\theta _k}} \right)} } \right)\left( {\sum\limits_{k = 1}^N {{\eta _j}\left( {{\theta _k}} \right)} } \right)
\end{align}
By introducing matrix $Y$ and assembling random samples, we have
\begin{equation}\label{Y}
  Y = \left[ {\left\{ {{\eta _1}\left( {{\theta _k}} \right)} \right\}_{k = 1}^N, \cdots ,\left\{ {{\eta _n}\left( {{\theta _k}} \right)} \right\}_{k = 1}^N} \right] = \left[ {\begin{array}{*{20}{c}} {{\eta _1}\left( {{\theta _1}} \right)}& \cdots &{{\eta _n}\left( {{\theta _1}} \right)}\\ \vdots &{}& \vdots \\ {{\eta _1}\left( {{\theta _N}} \right)}& \cdots &{{\eta _n}\left( {{\theta _N}} \right)} \end{array}} \right]
\end{equation}
Then Eq.\eqref{Tij_M} can be rewritten in matrix
\begin{equation}\label{T}
T = \frac{{{Y^T}Y}}{{N - 1}} - \frac{{{Y^T}U{U^T}Y}}{{N\left( {N - 1} \right)}}
\end{equation}
where $U = {\left[ 1 \right]_{N \times 1}}$, and $T$ is the simulated covariance matrix of random variables samples $\left\{ {\left\{ {{\eta _i}\left( {{\theta _k}} \right)} \right\}_{k = 1}^N} \right\}_{i = 1}^n$.

Both target covariance matrix $C$ and simulated covariance matrix $T$ are symmetric and positive definite, thus Cholesky decompositions can be used to performed on matrix $C$ and $T$ as
\begin{equation}\label{Cd}
  C= {P^T}P
\end{equation}
and
\begin{equation}\label{Td}
  T = {Q^T}Q
\end{equation}
where $Q$ and $P$ are upper triangular matrices.

In order to match the target covariance matrix $C$, a new random samples matrix $Y'$ is introduced as
\begin{equation}\label{Ynew}
  Y' = Y{Q^{ - 1}}P
\end{equation}
By the transformation in Eq.\eqref{Ynew}, simulated covariance matrix $T'$ can match the target covariance matrix $C$ and proof as follows,
\begin{proof}
	Simulated covariance matrix $T'$ of the new random samples $Y'$ is
	\begin{align*}
	  T' &= \frac{{{{Y'}^T}Y'}}{{N - 1}} - \frac{{{{Y'}^T}U{U^T}Y'}}{{N\left( {N - 1} \right)}} \\
	  & = \frac{{{P^T}{Q^{ - T}}{Y^T}Y{Q^{ - 1}}P}}{{N - 1}} - \frac{{{P^T}{Q^{ - T}}{Y^T}U{U^T}Y{Q^{ - 1}}P}}{{N\left( {N - 1} \right)}}\\
	  &= {P^T}{Q^{ - T}}\left( {\frac{{{Y^T}Y}}{{N - 1}} - \frac{{{Y^T}U{U^T}Y}}{{N\left( {N - 1} \right)}}} \right){Q^{ - 1}}P\\
	  &= {P^T}{Q^{ - T}}T{Q^{ - 1}}P \\
	  &= C
	\end{align*}
\end{proof}

It has to be noted that sample realizations $\left\{ {{\eta _i}\left( {{\theta _k}} \right)} \right\}_{k = 1}^N$ in each column of $Y'$ in Eq.\eqref{Ynew} are different from those in each column of $Y$ in Eq.\eqref{Y} due to the factor matrix $Q^{ - 1}P$. The realizations in each column of $Y'$ change with updated factor matrix $Q^{ - 1}P$ and do not match the marginal distribution function $F\left( {y;{x_i}} \right)$. 
A fact that re-ordering the sample realizations will not change distributions of random variables but change the statistical correlations, i.e., simulated covariance matrix, is enlightened. Hence, we use the strategy that re-ordering the sample realizations $\left\{ {{\eta _i}\left( {{\theta _k}} \right)} \right\}_{k = 1}^N$ in each column of $Y$  in Eq.\eqref{Y} follows the ranking of the realizations in each column of $Y'$ in Eq.\eqref{Ynew}. The target covariance matrix $C$ is matched by repeating the procedure of Eq.\eqref{T}, Eq.\eqref{Td}, Eq.\eqref{Ynew} and re-ordering stochastic samples.

The resulting procedure for simulating stochastic process samples is summarized in Algorithm \ref{Alg: Samples} as follows,
\begin{algorithm}[H]
\caption{Algorithm for simulating stochastic process samples }
\label{Alg: Samples}
\begin{algorithmic}[1]
\State Discretize spatial domain $x = \left\{ {{x_1}, \cdot  \cdot  \cdot ,{x_i}, \cdot  \cdot  \cdot ,{x_n}} \right\}$ and generate random variables samples $\left\{ {{\eta _i}\left({{\theta _k}} \right)} \right\}_{k = 1}^N\sim F\left( {y;{x_i}} \right), ~i = 1, \cdots, n$. \label{a11}
\State Compute the upper triangular matrix $P$ by use of a Cholesky decomposition in Eq.\eqref{Cd}. \label{a12}
\Repeat \label{a13}
    \State Compute upper triangular matrix ${Q^{\left( k-1 \right)}}$ by Eq.\eqref{Td} based on simulated covariance matrix ${T^{\left( k-1 \right)}}$ obtained by Eq.\eqref{T}. \label{a14}
    \State Compute $Y{'^{\left( k-1 \right)}}$ through Eq.\eqref{Ynew} and re-order samples in each column of $Y$ following the ranking of the realizations in each column of $Y'$ . \label{a15}
    \State Compute simulated covariance matrix ${T^{\left( k \right)}}$ by Eq.\eqref{T}. \label{a16}
    \Until ${{\left\| {{T^{\left( k \right)}} - C} \right\|} \mathord{\left/
 {\vphantom {{\left\| {{T^{\left( k \right)}} - C} \right\|} {\left\| C \right\|}}} \right.
 \kern-\nulldelimiterspace} {\left\| C \right\|}} < \varepsilon$ \label{a17}
\end{algorithmic}  
\end{algorithm}

Algorithm \ref{Alg: Samples} provides a simple and efficient framework to simulate stochastic processes samples. Spatial discrete points and initial random variables samples are generated in Step \ref{a11}. The number of random variables is equal to the that of spatial points and it's not necessary to discretize excessive spatial points. A Cholesky decomposition of target covariance matrix $C$ is performed in Step \ref{a12}. The computational cost can be neglected since only one time decomposition needs to be computed for matrix $C$. The Step \ref{a13} to Step \ref{a17} includes a loop iteration procedure to match the target covariance matrix $C$, and the computational cost in these steps is low since only Cholesky decompositions and re-ordering samples are involved. The convergence error in Step \ref{a17} can adopt 2-norm or infinite-norm (here same to 1-norm) and we adopt 2-norm in this paper. Note that, Algorithm \ref{Alg: Samples} can be applied to non-gaussian and non-stationary stochastic processes and can be readily generalized to high-dimensional random fields.

\section{Expansions of stochastic processes} \label{Sec3}
Algorithm \ref{Alg: Samples} provides an efficient procedure to simulate samples of stochastic processes. However, sample-based descriptions of stochastic processes are not suitable for subsequent applications in some cases \cite{ghanem2003stochastic, stefanou2009stochastic} and it's necessary to develop methods to represent the obtained samples of sochastic process. In general, a stochastic process $\omega \left( {x,\theta } \right)$ can be expressed as
\begin{equation}\label{exp}
\omega \left( {x,\theta } \right) = \sum\limits_{i = 0}^\infty  {{\xi _i}\left( \theta  \right){f_i}\left( x \right)}
\end{equation}
where $\left\{ {{\xi _i}\left( \theta  \right)} \right\}_{i = 0}^\infty$ and $\left\{ {{f_i}\left( x \right)} \right\}_{i = 0}^\infty$ are a set of random variables and deterministic functions, respectively. In practical, Eq.\eqref{exp} can be truncated at the term $M$ as
\begin{equation}\label{exp_t}
  \omega \left( {x,\theta } \right) = \sum\limits_{i = 0}^M {{\xi _i}\left( \theta  \right){f_i}\left( x \right)}
\end{equation}

There exist three unknown `variables' in Eq.\eqref{exp_t}, i.e., stochastic process $\omega \left( {x,\theta } \right)$, random variables $\left\{ {{\xi _i}\left( \theta  \right)} \right\}_{i = 0}^M$ and deterministic functions $\left\{ {{f_i}\left( x \right)} \right\}_{i = 0}^M$. Eq.\eqref{exp_t} can be determined if any two variables are available. 
In this context, only a set of random variables $\left\{ {{\xi _i}\left( \theta  \right)} \right\}_{i = 0}^M$ or deterministic functions $\left\{ {{f_i}\left( x \right)} \right\}_{i = 0}^M$ are required to be determined since samples of the stochastic process $\omega \left( {x,\theta } \right)$ have been obtained by Algorithm \ref{Alg: Samples}.
Three stratigies are developed to simulate stochastic processes through Eq.\eqref{exp_t}: \textit{(i).~select a set of deterministic basis $\left\{ {{f_i}\left( x \right)} \right\}_{i = 0}^M$, then compute random variables $\left\{ {{\xi _i}\left( \theta  \right)} \right\}_{i = 0}^M$; (ii).~select a set of random variables $\left\{ {{\xi _i}\left( \theta  \right)} \right\}_{i = 0}^M$, then compute deterministic functions $\left\{ {{f_i}\left( x \right)} \right\}_{i = 0}^M$; (iii).~select a set of deterministic basis $\left\{ {{f_i}\left( x \right)} \right\}_{i = 0}^M$ and random variables $\left\{ {{\xi _i}\left( \theta  \right)} \right\}_{i = 0}^M$, then compute unknown projection coefficients.} We'll develop corresponding algorithms based on above strategies in Section \ref{AKL}, Section \ref{APC} and Section \ref{AKLPC}, respectively.

\subsection{Algorithm based on stochastic samples and KL expansion} \label{AKL}
In order to obtain deterministic basis $\left\{ {{f_i}\left( x \right)} \right\}_{i = 0}^M$ in \textit{Strategy (i)}, we adopt Karhunen-Lo\`{e}ve (KL) expansion due to its minimum mean-square error property. The KL expansion is a special case of Eq.\eqref{exp} and is based on a spectral decomposition of the covariance function $C\left( {{x_1},{x_2}} \right)$ of the stochastic process $\omega \left( {x,\theta } \right)$ with the form
\begin{equation}\label{s1_1}
\omega \left( {x,\theta } \right) = \bar \omega \left( x \right) + \sum\limits_{i = 1}^M {{\xi _i}\left( \theta  \right)\sqrt {{\lambda _i}} {f_i}\left( x \right)}
\end{equation}
where $\bar \omega \left( x \right)$ is the mean function of the stochastic process $\omega \left( {x,\theta } \right)$, $M$ is the number of terms of KL expansion and $\left\{ {{\xi _i}\left( \theta  \right)} \right\}_{i = 1}^M$ is a set of uncorrelated random variables with zero mean and unit variance, i.e.,
\begin{equation}\label{s1_2}
  E\left\{ {{\xi _i}\left( \theta  \right)} \right\} = 0,~E\left\{ {{\xi _i}\left( \theta  \right){\xi _j}\left( \theta  \right)} \right\} = {\delta _{ij}}
\end{equation}
and given by
\begin{equation}\label{s1_3}
  {\xi _i}\left( \theta  \right) = \frac{1}{{\sqrt {{\lambda _i}} }}\int_D {\left[ {\omega \left( {x,\theta } \right) - \bar \omega \left( x \right)} \right]{f_i}\left( x \right)dx}
\end{equation}
where $\left\{ {{\lambda _i}} \right\}$ and $\left\{ {{f_i}\left( x \right)} \right\}$ are the eigenvalues and eigenfunctions of the covariance function $C\left( {{x_1},{x_2}} \right)$, obtained from solving the following homogeneous Fredholm integral equation of the second kind
\begin{equation}\label{s1_4}
  \int_D {C\left( {{x_1},{x_2}} \right){f_i}\left( {{x_1}} \right)d{x_1}}  = {\lambda _i}{f_i}\left( {{x_2}} \right)
\end{equation}
which satisfies
\begin{equation}\label{s1_5}
  \int_D {{f_i}\left( x \right){f_j}\left( x \right)dx}  = {\delta _{ij}}
\end{equation}

The solution of Eq.\eqref{s1_4} can be determined numerically for problems of practical interests. It is known that, for fixed $M$, the resulting random process approximation $\omega \left( {x,\theta } \right)$ is optimal among series expansion methods with respect to the global mean square error \cite{ghanem2003stochastic}. If the stochastic process $\omega \left( {x,\theta } \right)$ is gaussian, then $\left\{ {{\xi _i}\left( \theta  \right)} \right\}_{i = 1}^M$ are independent standard gaussian random variables. But for non-gaussian stochastic processes, $\left\{ {{\xi _i}\left( \theta  \right)} \right\}_{i = 1}^M$ are generally non-gaussian and there are no general methods for determining them. In order to determine their distributions, Eq.\eqref{s1_3} has to be solved. Here we adopt a sample-based method to compute the samples of random variables  $\left\{ {{\xi _i}\left( \theta  \right)} \right\}_{i = 1}^M$ as
\begin{equation}\label{s1_6}
  {\xi _i}\left( {{\theta _k}} \right) = \frac{1}{{\sqrt {{\lambda _i}} }}\int_D {\left[ {\omega \left( {x,{\theta _k}} \right) - \bar \omega \left( x \right)} \right]{f_i}\left( x \right)dx} ,~k = 1, \cdots ,N
\end{equation}
which needs $N$ times deterministic integral and has very low computational costs. The above method is summarized in Algorithm \ref{Alg: s1} as
\begin{algorithm}[H]
\caption{Algorithm based on stochastic samples and KL expansion}
\label{Alg: s1}
\begin{algorithmic}[1]
\State Generate samples of the stochastic process by use of Algorithm \ref{Alg: Samples}. \label{a21}
\State Solve a set of deterministic basis $\left\{ {{f_i}\left( x \right)} \right\}_{i = 1}^M$ by Eq.\eqref{s1_4}. \label{a22}
\State Compute random variables' samples $\left\{ {\left\{ {{\xi _i}\left( {{\theta _k}} \right)} \right\}_{k = 1}^N} \right\}_{i = 1}^M$ by Eq.\eqref{s1_6}. \label{a23}
\end{algorithmic}  
\end{algorithm}

Algorithm \ref{Alg: s1} provides an efficient method to expand stochastic samples ontained from Algorithm \ref{Alg: Samples}. Deterministic basis $\left\{ {{f_i}\left( x \right)} \right\}_{i = 1}^M$ are solved by use of KL eapansion, thus guarenting the minimum numbers of expansion items $M$. Random variables $\left\{ {{\xi _i}\left( \theta  \right)} \right\}_{i = 1}^M$ are described by samples and can be computed with very low computational costs. 

\subsection{Algorithm based on stochastic samples and PC expansion} \label{APC}
For the \textit{Strategy (ii)}, it's not easy to select random variables $\left\{ {{\xi _i}\left( \theta  \right)} \right\}_{i = 1}^M$  in Eq.\eqref{exp_t} directly, here we utilize Polynomial Chaos (PC) expansion to represent the random variables $\left\{ {{\xi _i}\left( \theta  \right)} \right\}_{i = 1}^M$ and then compute deterministic coefficient functions $\left\{ {{f_i}\left( x \right)} \right\}_{i = 1}^M$. A general representation of the stochastic process $\omega \left( {x,\theta } \right)$ in PC expansion has the following form
\begin{equation}\label{s2_1}
  \omega \left( {x,\theta } \right) = \bar \omega \left( x \right) + \sum\limits_{i = 1}^P {{\Gamma _i}\left( \theta  \right){f_i}\left( x \right)}
\end{equation}
where $\bar \omega \left( x \right)$ is the mean function of the stochastic proces $\omega \left( {x,\theta } \right)$, $\left\{ {{\Gamma _i}\left( \theta  \right)} \right\}_{i = 1}^P$ are Polynomial Chaos basis and can be generated by Rodriguez formula \cite{xiu2010numerical}. The deterministic coefficient functions $\left\{ {{f_i}\left( x \right)} \right\}_{i = 1}^M$ can be obtained by
\begin{equation}\label{s2_2}
  {f_i}\left( x \right) = \frac{{E\left\{ {\left[ {\omega \left( {x,\theta } \right) - \bar \omega \left( x \right)} \right]{\Gamma _i}\left( \theta  \right)} \right\}}}{{E\left\{ {\Gamma _i^2\left( \theta  \right)} \right\}}}
\end{equation}
where $E\left\{  \cdot  \right\}$ is the expectation operator and the above method is summarized in Algorithm \ref{Alg: s2} as
\begin{algorithm}[H]
\caption{Algorithm based on stochastic samples and PC expansion}
\label{Alg: s2}
\begin{algorithmic}[1]
\State Generate samples of the stochastic process by use of Algorithm \ref{Alg: Samples}. \label{a31}
\State Choose standard random variables $\left\{ {{\gamma _i}\left( \theta  \right)} \right\}_{i = 1}^{{M_\gamma }}$. \label{a32}
 \State Generate PC basis $\left\{ {{\Gamma _i}\left( \theta  \right)} \right\}_{i = 1}^P$ of random variables $\left\{ {{\gamma _i}\left( \theta  \right)} \right\}_{i = 1}^{{M_\gamma }}$. \label{a33}
\State Compute $\left\{ {{f_i}\left( x \right)} \right\}_{i = 1}^P$ by Eq.\eqref{s2_2}. \label{a34}
\end{algorithmic}  
\end{algorithm}

There are various choices for specifying random variables ${\gamma _i}\left( \theta  \right)$ in Step \ref{a32}, such as gaussian random variables and uniform random variables, and the corrsponding PC basis should be adopted as Hermite Polynomial Chaos basis \cite{ghanem2003stochastic} and Generalized Polynomial Chaos basis \cite{ghanem2003stochastic, xiu2010numerical}, respectively. Further, Algorithm \ref{Alg: s2} provides a suitable method to simulate stochastic processes if Eq.\eqref{s1_4} is not easy to solve. However the computational efficiency decreases if the number $M_\gamma$ of random variables ${\gamma _i}\left( \theta  \right)$ or the order of PC basis is too large.

\subsection{Algorithm based on stochastic samples and KL-PC expansion} \label{AKLPC}
A natural idea is to combine methods in Section \ref{AKL} and Section \ref{APC} for the \textit{Strategy (iii)}. Hence, we choose deterministic functions $\left\{ {{f_i}\left( x \right)} \right\}_{i = 1}^M$ obtained by Eq.\eqref{s1_4} and express random variable ${\xi _i}\left( \theta  \right)$ in Eq.\eqref{s1_1} by use of PC expansion as
\begin{equation}\label{s3_1}
  {\xi _i}\left( \theta  \right) = \sum\limits_{j = 1}^P {{c_{ij}}{\Gamma _j}\left( \theta  \right)}
\end{equation}
Substituting Eq.\eqref{s3_1} into Eq.\eqref{s1_1} yields
\begin{equation}\label{s3_2}
\omega \left( {x,\theta } \right) = \bar \omega \left( x \right) + \sum\limits_{i = 1}^M {\sum\limits_{j = 1}^P {{c_{ij}}{\Gamma _j}\left( \theta  \right)\sqrt {{\lambda _i}} {f_i}\left( x \right)} }
\end{equation}
Hence, unknown projection coefficients $c_{ij}$ can be computed by
\begin{equation}\label{s3_3}
{c_{ij}} = \frac{1}{{\sqrt {{\lambda _i}} E\left\{ {\Gamma _j^2\left( \theta  \right)} \right\}}}\int_D {E\left\{ {\left[ {\omega \left( {x,\theta } \right) - \bar \omega \left( x \right)} \right]{\Gamma _j}\left( \theta  \right)} \right\}{f_i}\left( x \right)dx}
\end{equation}
The above method is summarized in Algorithm \ref{Alg: s3} as
\begin{algorithm}[H]
\caption{Algorithm based on stochastic samples and KL-PC expansion}
\label{Alg: s3}
\begin{algorithmic}[1]
\State Generate samples of the stochastic process by use of Algorithm \ref{Alg: Samples}. \label{a41}
\State Solve a set of deterministic basis $\left\{ {{f_i}\left( x \right)} \right\}_{i = 1}^M$ by Eq.\eqref{s1_4}. \label{a42}
\State Choose standard random variables $\left\{ {{\gamma _i}\left( \theta  \right)} \right\}_{i = 1}^{{M_\gamma }}$. \label{a43}
 \State Generate PC basis $\left\{ {{\Gamma _i}\left( \theta  \right)} \right\}_{i = 1}^P$ of random variables $\left\{ {{\gamma _i}\left( \theta  \right)} \right\}_{i = 1}^{{M_\gamma }}$.  \label{a44}
\State Compute ${\left\{ {{c_{ij}}} \right\}_{i,j = 1}}$ by Eq.\eqref{s3_3}. \label{a45}
\end{algorithmic}  
\end{algorithm}

Algorithm \ref{Alg: s3} combines KL expansion and PC expansion since PC expansion provides an explicit representation for random variables $\left\{ {{\xi _i}\left( \theta  \right)} \right\}_{i = 1}^M$ obtained in Algorithm \ref{Alg: s1}. Algorithm \ref{Alg: s3} is a good choice if sample-based descriptions are not convenient for some practical problems. Further, stochastic processes may be described directly by real samples instead of being specified by their covariance functions and marginal probability distributions. We can use real samples instead of stochastic samples in Step \ref{a41} of Algorithm \ref{Alg: s1}, Algorithm \ref{Alg: s2} and Algorithm \ref{Alg: s3}, thus Algorithm \ref{Alg: Samples} is no longer necessary.

This section provides three numerical algorithms to expand the stochastic samples obtained by Algorithm \ref{Alg: Samples}. Different algorithms can be chosen according to properties of practical problems and the performances of proposed algorithms are shown in next section.

\section{Numerical examples} \label{Sec4}
Three numerical examples, including a non-gaussian and stationary stochastic process, a non-gaussian and non-stationary stochastic process and a strongly non-gaussian and non-stationary stochastic process,  are used to demonstrate the accuracies and efficiencies of the proposed algorithms. The performances of Algorithm \ref{Alg: s1}, Algorithm \ref{Alg: s2} and Algorithm \ref{Alg: s3} are indicated in Example \ref{Example1}, and only Algorithm \ref{Alg: s1} is used to represent stochastic samples in Example \ref{Example2} and Example \ref{Example3}. Since the simulated marginal distribution function automatically matches the target one, the accuracies of the proposed methods are validated by comparing the simulated covariance function with the exact one. For all examples, spatial domains are discretized as 50 points, convergence errors are set as $5 \times {10^{ - 3}}$  and $1 \times {10^{ 4}}$ initial samples for each random variable  are generated.

\subsection{Example 1: non-gaussian and stationary stochastic process} \label{Example1}
Consider a stochastic process with beat marginal distribution function \cite{phoon2002simulation}
\begin{equation}\label{e1_1}
F\left( {y;p,q} \right) = \frac{{\Gamma \left( {p + q} \right)}}{{\Gamma \left( p \right)\Gamma \left( q \right)}}\int_0^u {{z^{p - 1}}{{\left( {1 - z} \right)}^{q - 1}}dz}
\end{equation}
and squared exponential covariance function
\begin{equation}\label{e1_2}
C\left( {{x_1},{x_2}} \right) = {e^{ - {{\left( {{x_1} - {x_2}} \right)}^2}}}
\end{equation}
In Eq.\eqref{e1_1}, $\Gamma \left(  \cdot  \right)$ is the gamma function and parameters are given by 
\begin{equation}\label{e1_3}
u = \frac{{y - {y_{\min }}}}{{{y_{\max }} - {y_{\min }}}}, ~p = 4, ~q = 2
\end{equation}

The expectation function and variance function of the beat distribution in Eq.\eqref{e1_1} are
\begin{equation}\label{e1_4}
\left\{ \begin{array}{l}
{\mu _F}\left( x \right) = \left( {{y_{\max }} - {y_{\min }}} \right)\frac{p}{{p + q}} + {y_{\min }}\\
\sigma _F^2\left( x \right) = {\left( {{y_{\max }} - {y_{\min }}} \right)^2}\frac{{pq}}{{{{\left( {p + q} \right)}^2}\left( {p + q + 1} \right)}}
\end{array} \right.
\end{equation}
According to Eq.\eqref{e1_2}, we can obtain $\sigma _F^2\left( x \right) = 1$. Letting ${\mu _F}\left( x \right) = 0$ and solving Eq.\eqref{e1_4} yield ${y_{\min }} =  - \sqrt {14}  \approx  - 3.74$, ${y_{\max }} = \sqrt {3.5} \approx 1.87$.

\begin{figure}[H]
	\begin{center}
		\includegraphics[width=0.8\textwidth]{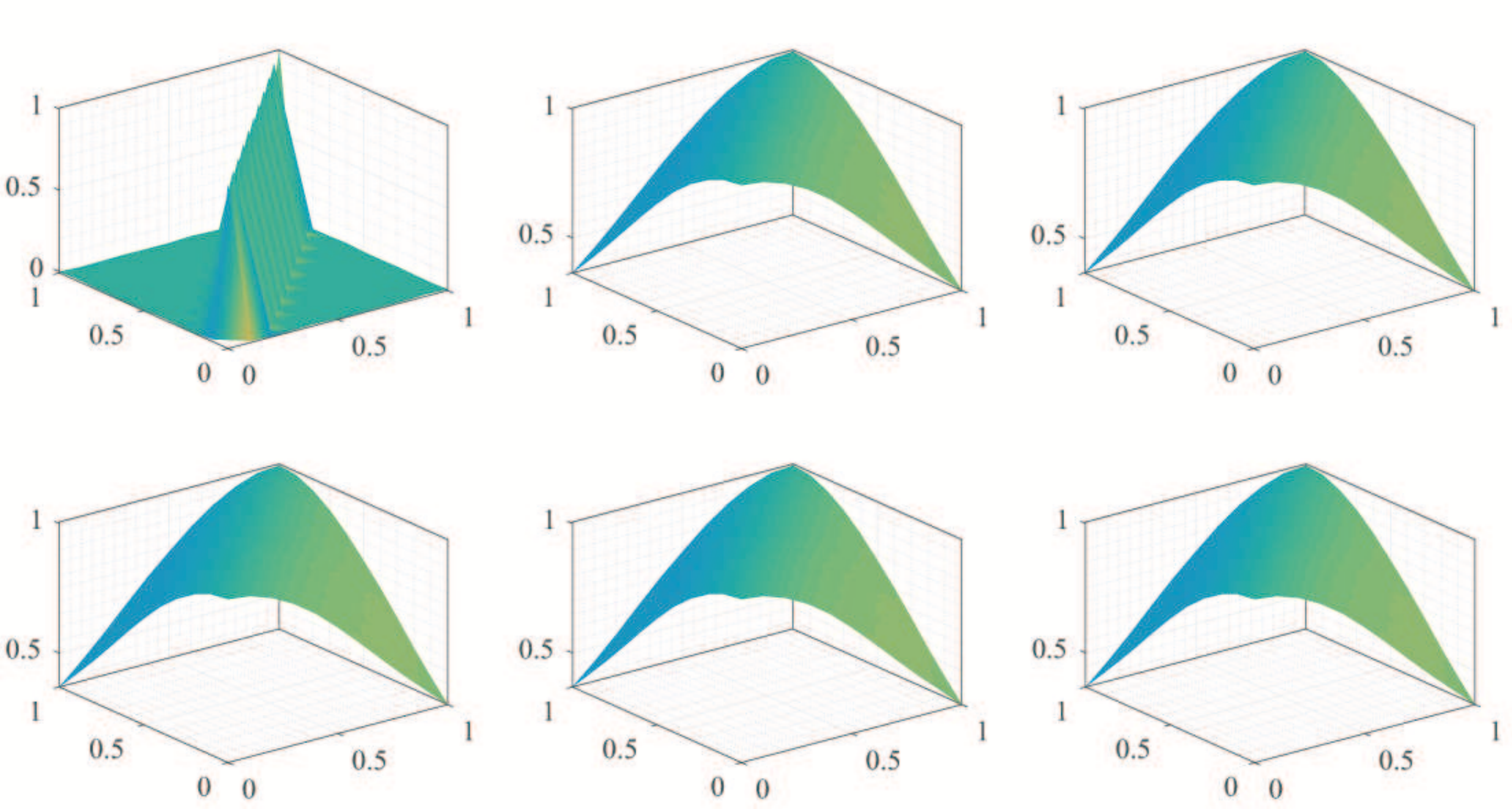} 
		\caption{Iterations of simulated covariance functions $\left\{ {{T^{\left( k \right)}}} \right\}_{k = 0}^5$.}
		\label{SP101}
	\end{center}
\end{figure}

\begin{figure}[H]
	\begin{center}
		\includegraphics[width=0.7\textwidth]{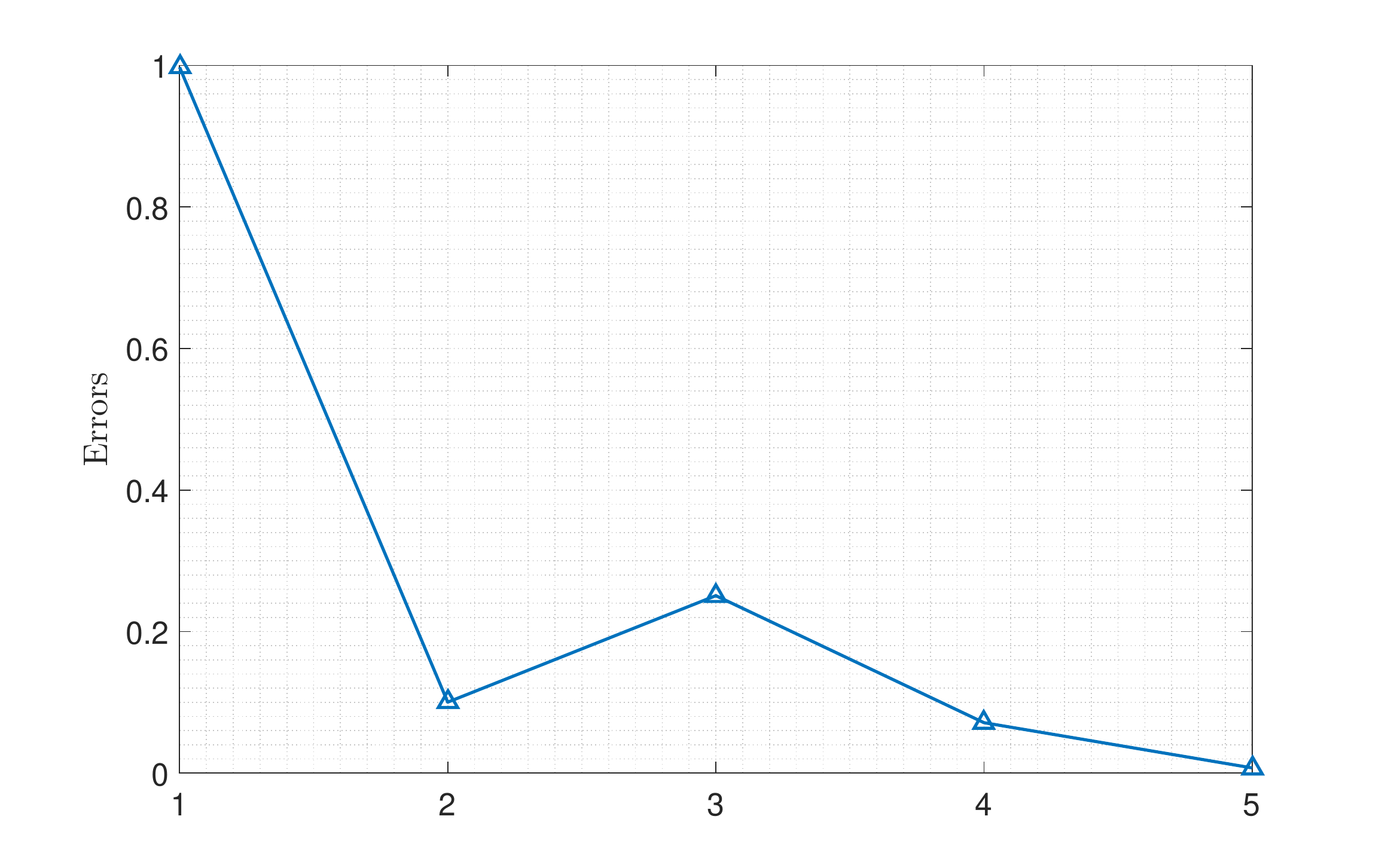}
		\caption{Iterative errors.}
		\label{SP102}
	\end{center}
\end{figure}

Fig.\ref{SP101} shows iterations of sample covariance functions $\left\{ {{T^{\left( k \right)}}} \right\}_{k = 0}^5$, where ${T^{\left( 0 \right)}}$ is the covariance function of the initial stochastic samples. It indicates that a rough approximation of the target covariance function can be obtained with only one round of iteration. The convergence error in each iteration is shown in Fig.\ref{SP102}, which demonstrates Algorithm \ref{Alg: Samples} has a good convergence and a high accuracy.

\begin{figure}[H]
	\begin{center}
		\includegraphics[width=1.0\textwidth]{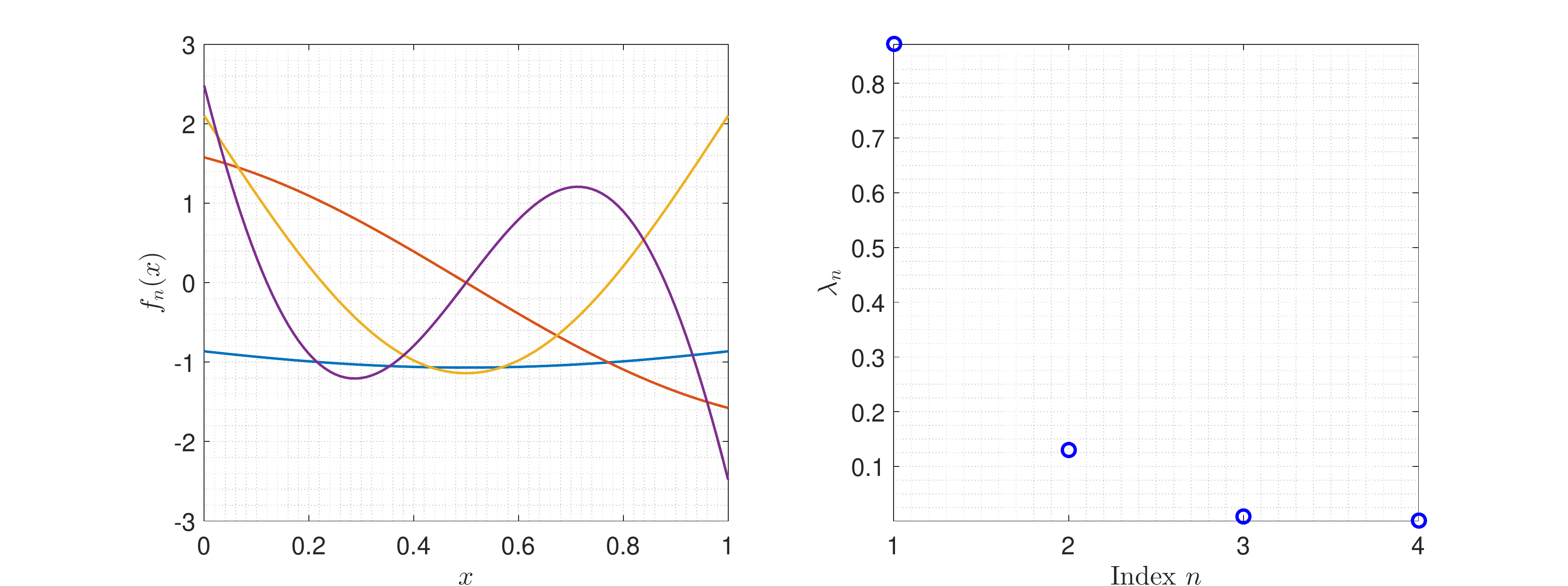}
		\caption{Eigenfunctions (left) and eigenvalues (right) of covariance $C\left( {{x_1},{x_2}} \right)$.}
		\label{SP103}
	\end{center}
\end{figure}

In order to verify the performances of the porposed methods, we utilize Algorithm \ref{Alg: s1}, Algorithm \ref{Alg: s2} and Algorithm \ref{Alg: s3} to represent obtained stochastic samples, respectively. For Algorithm \ref{Alg: s2}, Fig.\ref{SP103} shows first four eigenfunctions and eigenvalues of $C\left( {{x_1},{x_2}} \right)$ obtained by Eq.\eqref{s1_4}.

\begin{figure}[H]
	\begin{center}
		\includegraphics[width=0.7\textwidth]{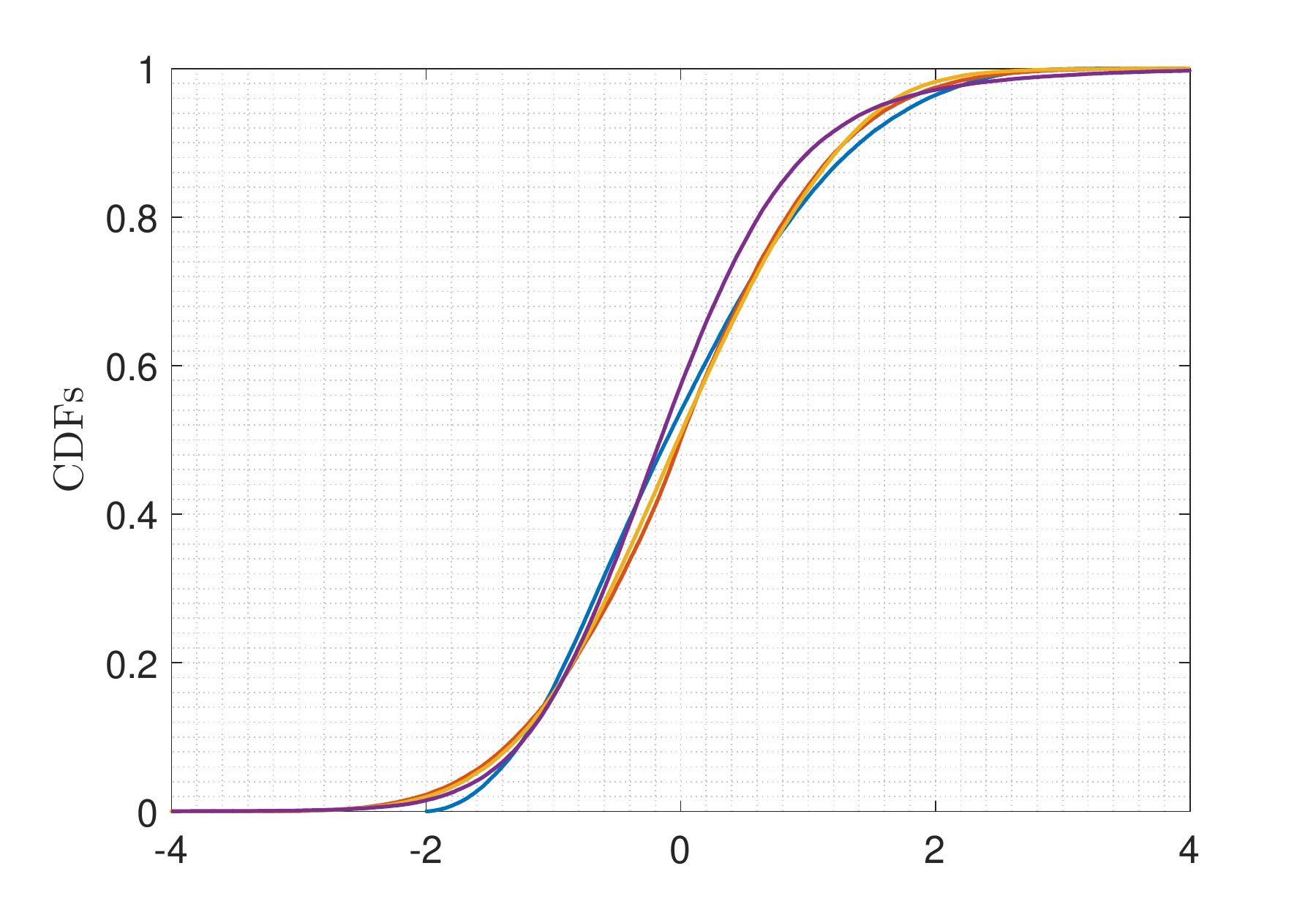}
		\caption{CDFs of random variables $\left\{ {{\xi _i}\left( \theta  \right)} \right\}_{i = 1}^4$.}
		\label{SP104}
	\end{center}
\end{figure}

Fig.\ref{SP104} shows cumulative distribution functions (CDFs) of random variables $\left\{ {{\xi _i}\left( \theta  \right)} \right\}_{i = 1}^4$ computed by Eq.\eqref{s1_6}
and Table.\ref{E1t1} indicates that they are uncorrelated random variables, which is consistent with the theory of KL expansion.
\begin{table}[H]
	\centering	
	\caption{Statistical correlations between ${\xi _i}\left( \theta  \right)$ and ${\xi _j}\left( \theta  \right)$}\vspace{-0.6em}
	\label{E1t1}
	\begin{tabular}{rrrrr}      
	      \toprule 
		$E\left\{ {{\xi _i}{\xi _j}} \right\}$ & ${\xi _1}\left( \theta  \right)$ & ${\xi _2}\left( \theta  \right)$ & ${\xi _3}\left( \theta  \right)$ & ${\xi _4}\left( \theta  \right)$ \\
		\midrule 
		${\xi _1}\left( \theta  \right)$ &  1.0005 \\
		${\xi _2}\left( \theta  \right)$ &  0.0004  &  0.9995  &  & sym.\\
		${\xi _3}\left( \theta  \right)$ &  0.0042  &  0.0016  & 0.9927  \\
		${\xi _4}\left( \theta  \right)$ &  0.0272  &  0.0087  & $-$0.0082 &  1.0248 \\
		\bottomrule 
	\end{tabular}
\end{table}

\begin{figure}[H]
	\begin{center}
		\includegraphics[width=0.8\textwidth]{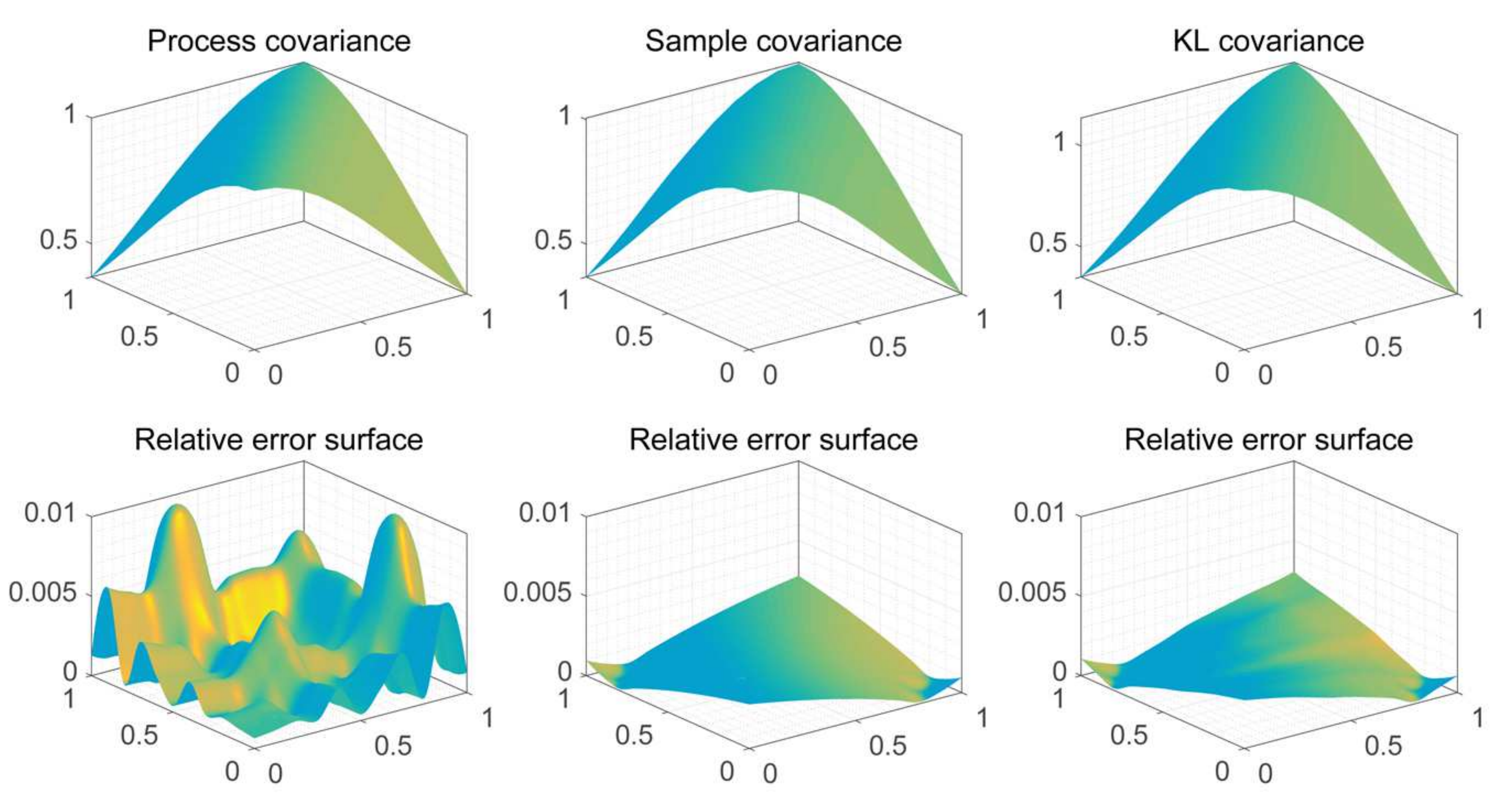}
		\caption{Comparisons between target, sample and KL-simulated covariance.}
		\label{SP105}
	\end{center}
\end{figure}

Fig.\ref{SP105} shows comparisons between target (top left), sample (top mid) and KL-simulated (top right) covariance. The relative errors between target and sample covariance (bottom left), target and KL-simulated covariance (bottom mid) and sample and KL-simulated covariance (bottom right) demonstrate the high accuracy of the proposed Algorithm \ref{Alg: s1}.

\begin{figure}[H]
	\begin{center}
		\includegraphics[width=0.7\textwidth]{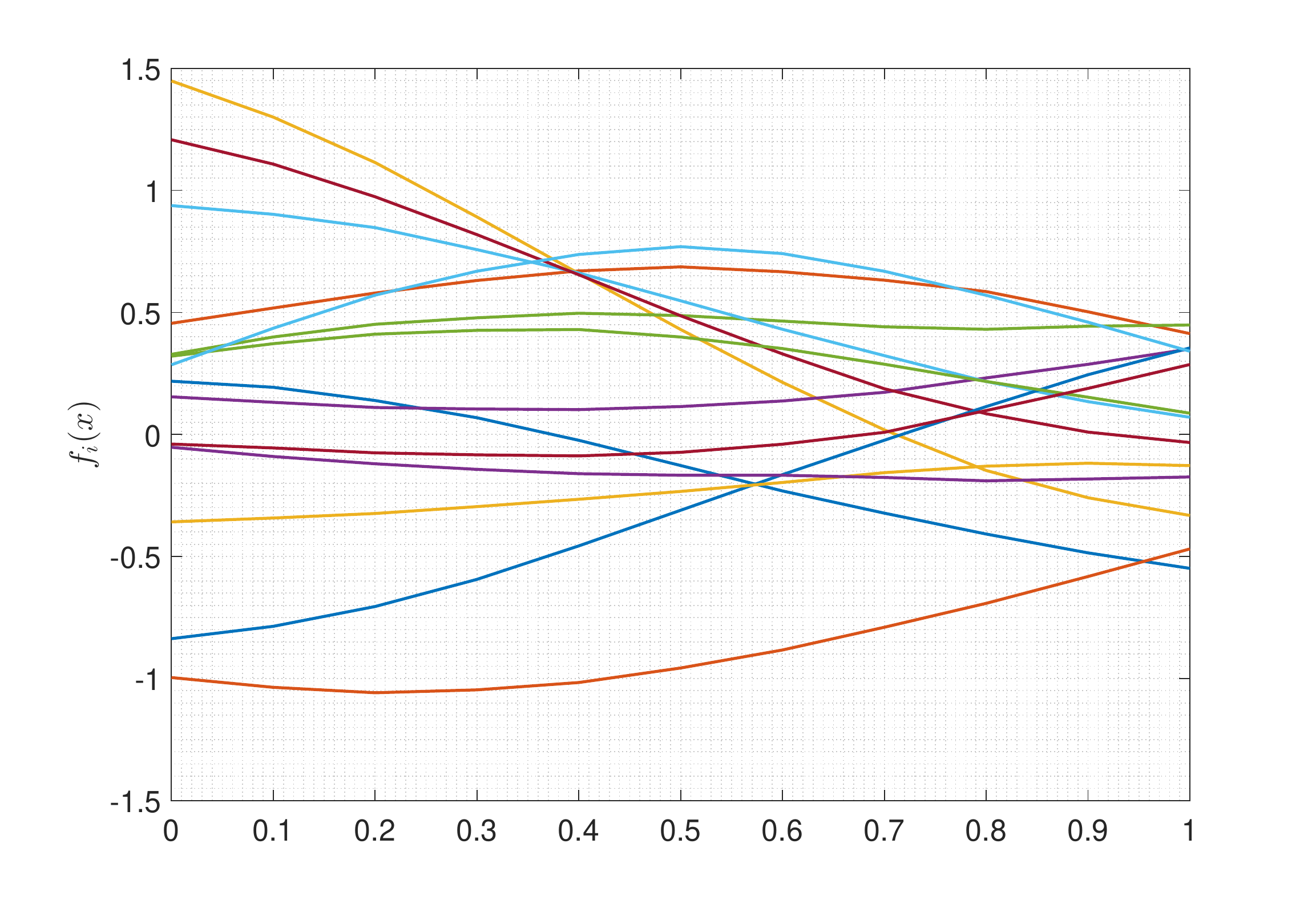}
		\caption{Coefficient functions $\left\{ {{f_i}\left( x \right)} \right\}_{i = }^{14}$ of PC expansion.}
		\label{SP106}
	\end{center}
\end{figure}

For Algorithm \ref{Alg: s2}, we adopt 2-order Hermite Polynomial Chaos basis, the number of total terms $M$ is 14. Fig.\ref{SP106} shows $\left\{ {{f_i}\left( x \right)} \right\}_{i = 1}^P$ obtained by Eq.\eqref{s2_2}.

\begin{figure}[H]
	\begin{center}
		\includegraphics[width=0.8\textwidth]{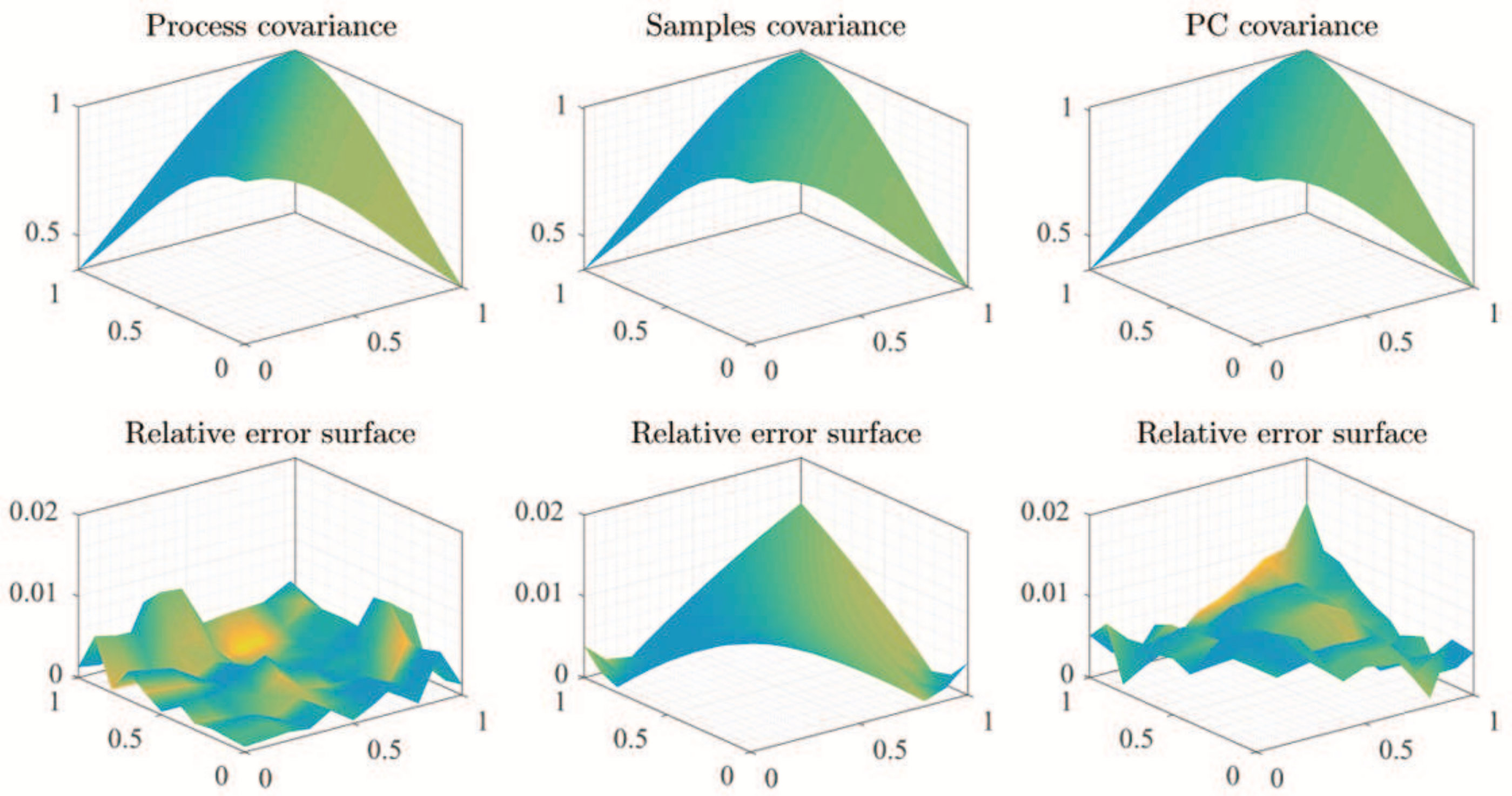}
		\caption{Comparisons between target, sample and PC-simulated covariance.}
		\label{SP107}
	\end{center}
\end{figure}

Comparisons between target, sample and PC-simulated covariance (the corresponding legends of relative errors are the same as Fig.\ref{SP105}) are shown in Fig.\ref{SP107}, which demonstrates the high accuracy of the proposed Algorithm \ref{Alg: s2}.

\begin{figure}[H]
	\begin{center}
		\includegraphics[width=0.8\textwidth]{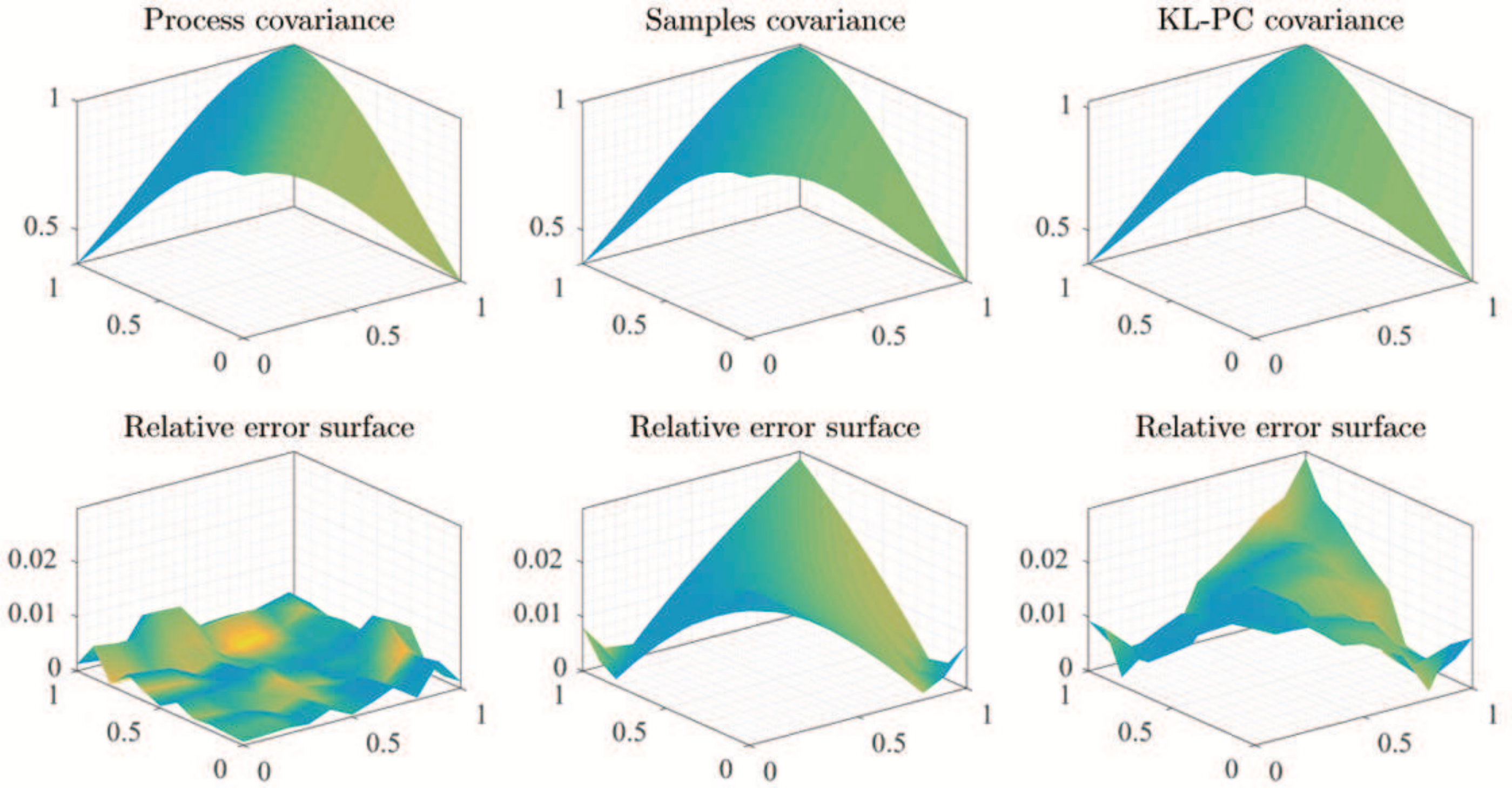}
		\caption{Comparison between target, sample and KL-PC-simulated covariance.}
		\label{SP108}
	\end{center}
\end{figure}

For Algorithm \ref{Alg: s3}, Table.\ref{E1t2} shows projection coefficients $c_{ij}$ obtained by Eq.\eqref{s3_3} and comparisons between target, sample and KL-PC-simulated covariance (the corresponding legends of relative errors are the same as Fig.\ref{SP105}) are shown as Fig.\ref{SP108}, which demonstrates the high accuracy of the proposed Algorithm \ref{Alg: s3}.
\begin{table}[H]
     \centering
	\caption{Coefficients $c_{ij}$ of the KL-PC expansion}\vspace{-0.6em}
	\label{E1t2}
	\begin{tabular}{rrrrrrrr}
		\toprule
		$c_{ij}$ & 1 & 2 & 3 & 4 & 5 & 6 & 7 \\
		\midrule
		1  &$-$0.2273	&$-$0.3286	&$-$0.6603	&$-$0.4586	&0.0511	&0.5482	&0.3417 \\
		2  &$-$0.3109	&0.8586	&0.2774	&0.5514	&$-$0.9518	&$-$0.6500	&$-$0.6917 \\
		3  &0.3333&$-$0.0232&1.1992	&$-$0.1781	&0.3611	&0.0587	&$-$0.3919 \\
		4  &$-$0.3514&$-$1.0478&0.7115	&$-$0.1398	&1.4399	&1.0510	&0.5405 \\
		\toprule
		$c_{ij}$ & 8 & 9 & 10 & 11 & 12 & 13 & 14 \\
		\midrule
		1  &0.2638	&$-$0.0346	&$-$0.0014	&0.3181	&$-$0.3325	&$-$0.5259	&0.7197 \\
		2  &0.4569	&0.0296	&0.7990	&$-$0.3317	&$-$0.2799	&$-$0.0816&0.2825 \\
		3  &$-$1.0618	&0.7456	&0.3166	&$-$0.6234	&0.2455	&$-$0.1098	&$-$0.4125 \\
		4  &$-$1.3680	&$-$0.2531	&$-$0.2757	&$-$0.3412	&0.0206	&$-$0.1233 &$-$0.9378 \\
		\bottomrule
	\end{tabular}
\end{table}

\subsection{Example 2: non-gaussian and non-stationary stochastic process} \label{Example2}
Consider the stochastic process with beat marginal distribution function in Eq.\eqref{e1_1} and Brown-Bridge covariance function \cite{phoon2002simulation}
\begin{equation}\label{e2_1}
C\left( {{x_1},{x_2}} \right) = \min \left( {{x_1},{x_2}} \right) - {x_1}{x_2}
\end{equation}
According to Eq.\eqref{e2_1}, we can obtain $\sigma _F^2\left( x \right) = C\left( {x,x} \right) = x - {x^2}$. Letting ${\mu _F}\left( x \right) = 0$ and solving Eq.\eqref{e1_4} yields,
\begin{equation}\label{e2_2}
{y_{\min }} =  - \sqrt {14\left( {x - {x^2}} \right)} , ~{y_{\max }} = \sqrt {3.5\left( {x - {x^2}} \right)}
\end{equation}

\begin{figure}[H]
	\begin{center}
		\includegraphics[width=1.0\textwidth]{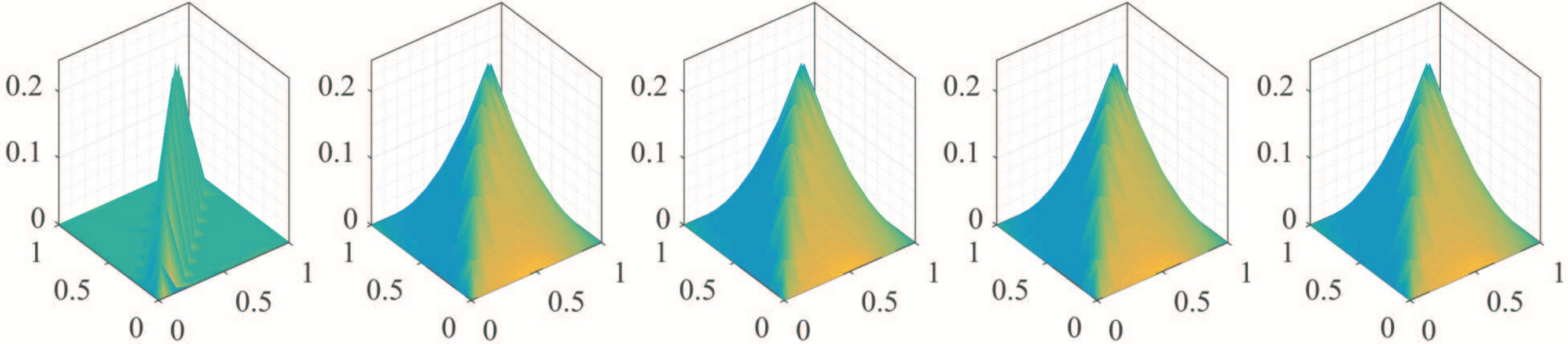}
		\caption{Iterations of sample covariance functions $\left\{ {{T^{\left( k \right)}}} \right\}_{k = 0}^4$.}
		\label{SP201}
	\end{center}
\end{figure}

\begin{figure}[H]
	\begin{center}
		\includegraphics[width=0.7\textwidth]{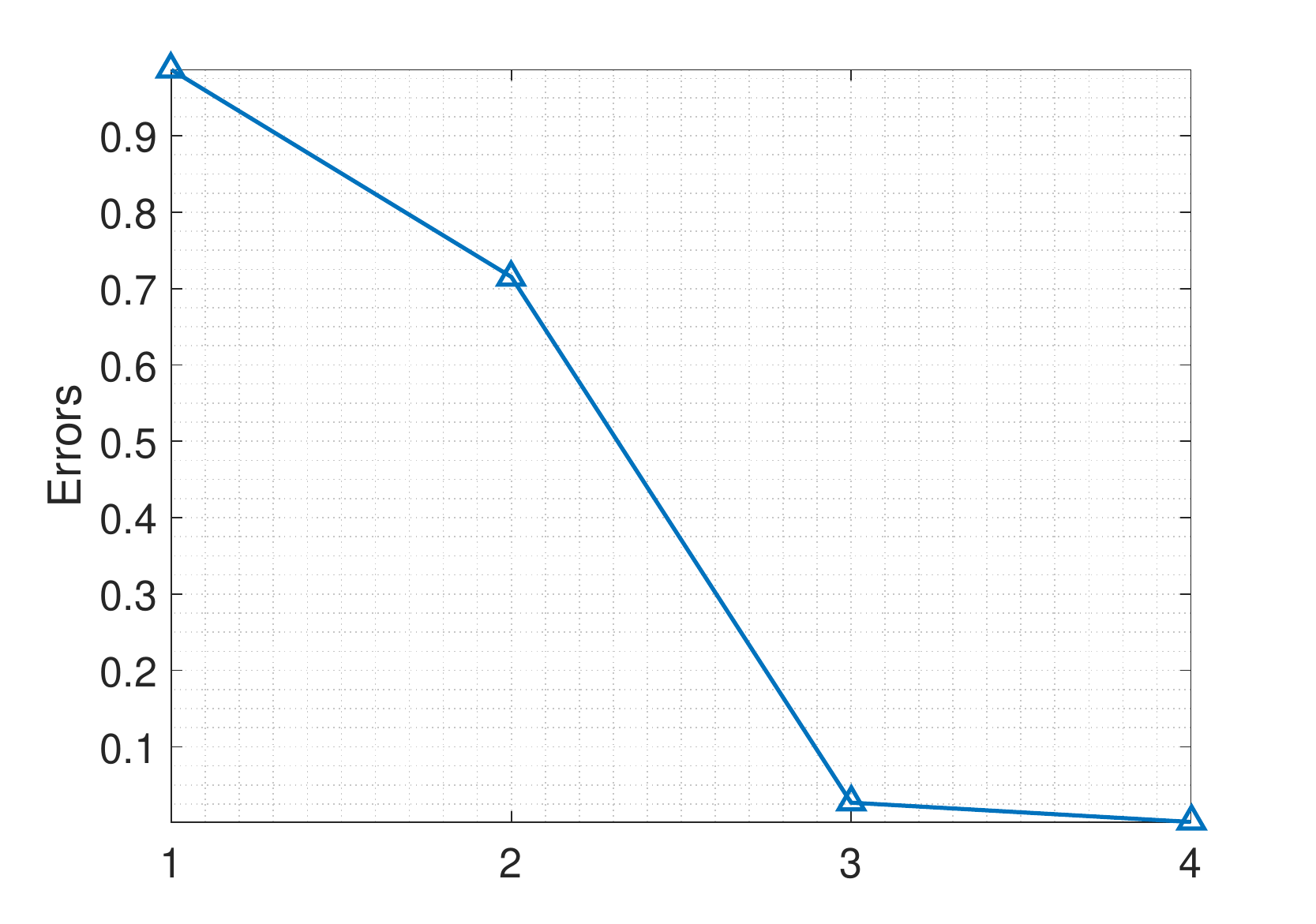}
		\caption{Iterative errors.}
		\label{SP202}
	\end{center}
\end{figure}

Fig.\ref{SP201} shows iterations of sample covariance functions $\left\{ {{T^{\left( k \right)}}} \right\}_{k = 0}^4$ and Fig.\ref{SP202} shows convergence error in each iteration, which once again demonstrate the good convergence and the high accuracy of Algorithm \ref{Alg: Samples}.

\begin{figure}[H]
	\begin{center}
		\includegraphics[width=1.0\textwidth]{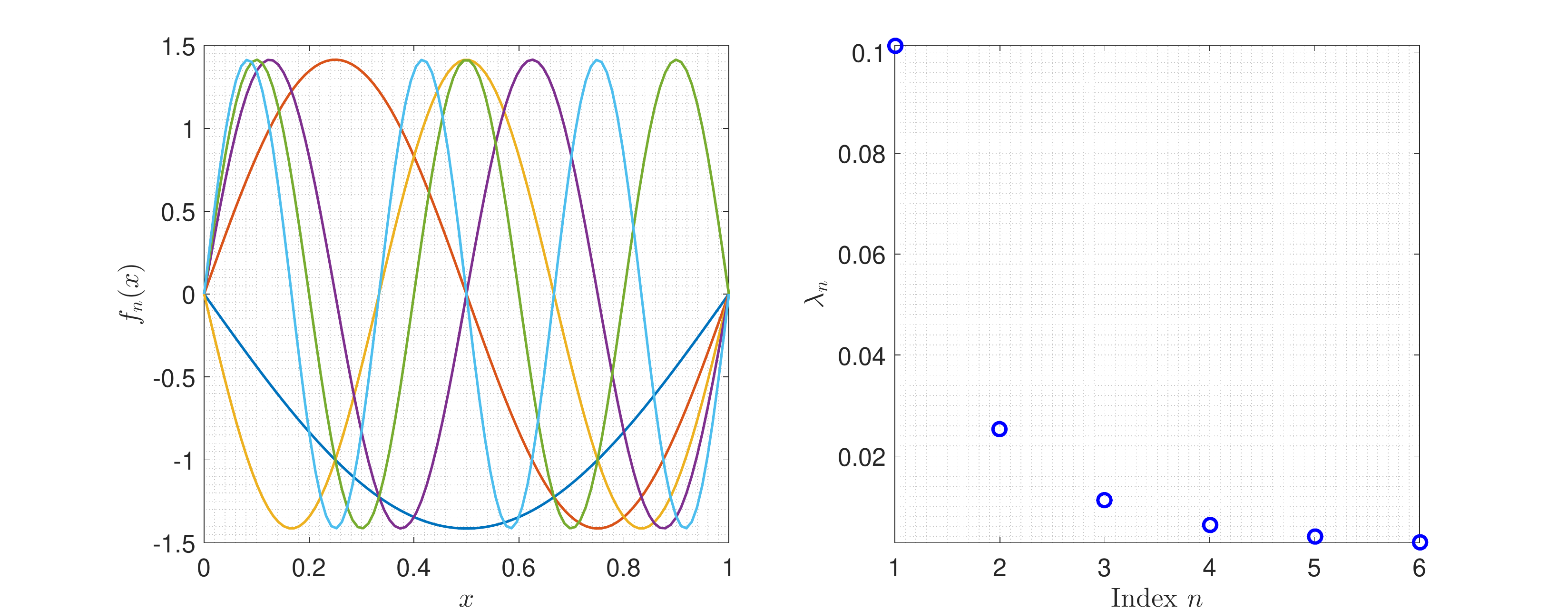}
		\caption{Eigenfunctions (left) and eigenvalues (right) of covariance $C\left( {{x_1},{x_2}} \right)$.}
		\label{SP203}
	\end{center}
\end{figure}

Example \ref{Example1} validate Algorithm \ref{Alg: s1}, Algorithm \ref{Alg: s2} and Algorithm \ref{Alg: s3}, here we only consider using Algorithm \ref{Alg: s1} expand the obtained stochastic samples from Algorithm \ref{Alg: Samples}. 
According to Eq.\eqref{s1_4}, analytical eigenfunctions and eigenvalues of $C\left( {{x_1},{x_2}} \right)$ can be obtained as
\begin{equation}\label{e2_3}
{f_k}\left( x \right) = \sqrt 2 \sin k\pi x, ~{\lambda}_k  = \frac{1}{{{k^2}{\pi ^2}}},~k = 1, 2, \cdots
\end{equation}
First six eigenfunctions and eigenvalues are shown in Fig.\ref{SP203}.

\begin{figure}[H]
	\begin{center}
		\includegraphics[width=0.7\textwidth]{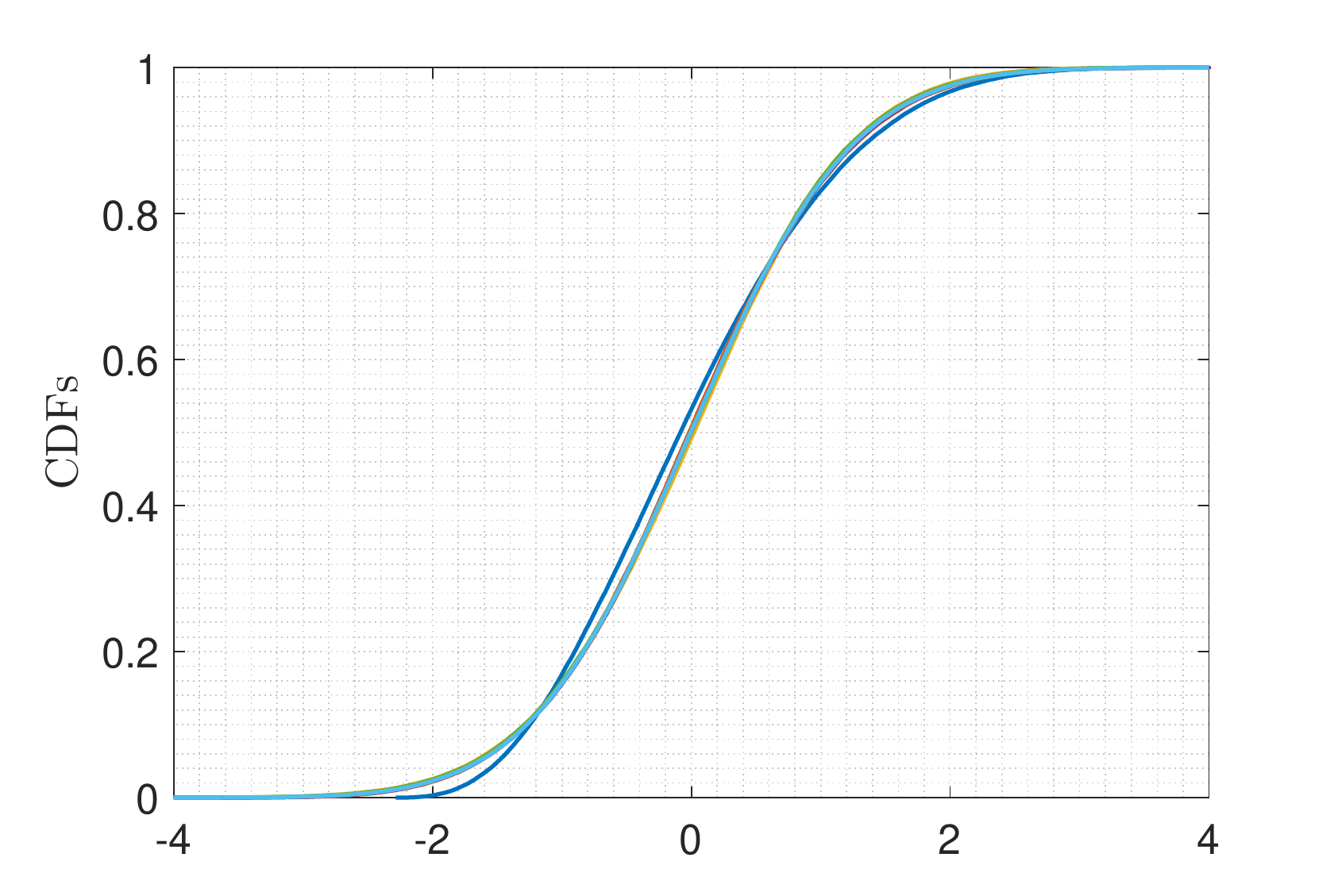}
		\caption{CDFs of random variables $\left\{ {{\xi _i}\left( \theta  \right)} \right\}_{i = 1}^6$.}
		\label{SP204}
	\end{center}
\end{figure}

\begin{figure}[H]
	\begin{center}
		\includegraphics[width=0.8\textwidth]{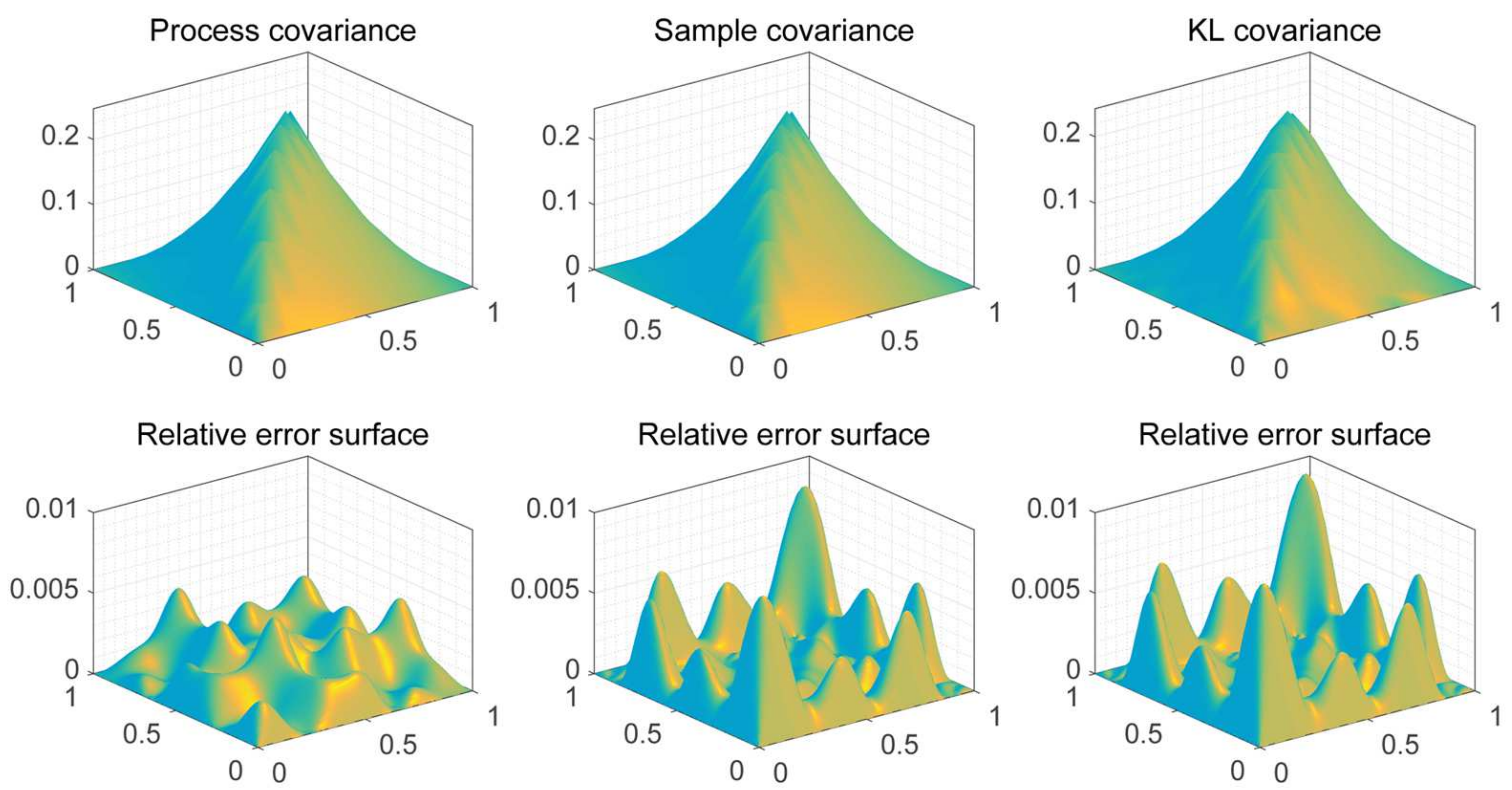}
		\caption{Comparisons between target, sample and KL-simulated covariance.}
		\label{SP205}
	\end{center}
\end{figure}

Fig.\ref{SP204} shows CDFs of random variables $\left\{ {{\xi _i}\left( \theta  \right)} \right\}_{i = 1}^6$ and they are uncorrelated as shown as Table.\ref{E2t1}, which is once again consistent with the theory of KL expansion. Fig.\ref{SP205} shows comparisons between target, sample and KL-simulated covariance (the corresponding legends of relative errors are the same as Fig.\ref{SP105}), which verify the applicability of the proposed Algorithm \ref{Alg: s1} to non-gaussian and non-stationary stochastic processes.
\begin{table}[H]
     \centering
	\caption{Statistical correlations between ${\xi _i}\left( \theta  \right)$ and ${\xi _j}\left( \theta  \right)$}\vspace{-0.6em}
	\label{E2t1}
	\begin{tabular}{rrrrrrr}
		\toprule
		$E\left\{ {{\xi _i}{\xi _j}} \right\}$ & ${\xi _1}\left( \theta  \right)$ & ${\xi _2}\left( \theta  \right)$ & ${\xi _3}\left( \theta  \right)$ & ${\xi _4}\left( \theta  \right)$ & ${\xi _5}\left( \theta  \right)$ & ${\xi _6}\left( \theta  \right)$ \\
		\midrule
		${\xi _1}\left( \theta  \right)$ &  0.9991 \\
		${\xi _2}\left( \theta  \right)$ &  $-$0.0010  &  1.0006 \\
		${\xi _3}\left( \theta  \right)$ &  $-$0.0025  &  0.0015  & 1.0005    &  & sym. \\
		${\xi _4}\left( \theta  \right)$ &  $-$0.0011  &  0.0017  & 0.0004 &  1.0001 \\
		${\xi _5}\left( \theta  \right)$ &  $-$0.0004  &  0.0006  & $-$0.0010 &  $-$0.0006  &  1.0000 \\
		${\xi _6}\left( \theta  \right)$ &  0.0026   &  $-$0.0009  & $-$0.0003 &   0.0009 &  0.0005 &  0.9999 \\
		\bottomrule
	\end{tabular}
\end{table}

\subsection{Example 3: strongly non-gaussian and non-stationary stochastic process} \label{Example3}
Consider a stochastic process with shifted lognormal marginal distribution proposed in \cite{phoon2002simulation}
\begin{equation}\label{e3_1}
F\left( {y;\mu ,\delta ,\sigma } \right) = \Phi \left( {\frac{{\ln \left( {y - \delta \left( x \right)} \right) - \mu \left( x \right)}}{\sigma }} \right)
\end{equation}
and covariance function
\begin{equation}\label{e3_2}
C\left( {{x_1},{x_2}} \right) = e^{ - \left( {{x_1} + {x_2}} \right) - \left| {{x_1} - {x_2}} \right|}
\end{equation}

The expectation function and variance function of the shifted lognormal distribution in Eq.\eqref{e3_1} are
\begin{equation}\label{e3_3}
\left\{ \begin{array}{l}
{\mu _F}\left( x \right) = \delta \left( x \right) + {e^{\mu \left( x \right) + \frac{1}{2}{\sigma ^2}}}\\
\sigma _F^2\left( x \right) = \left( {{e^{{\sigma ^2}}} - 1} \right){e^{2\mu \left( x \right) + {\sigma ^2}}}
\end{array} \right.
\end{equation}
According to Eq.\eqref{e3_2}, the variance function is $\sigma _F^2\left( x \right) = C\left( {x,x} \right) = e^{ - 2x}$. Letting ${\mu _F}\left( x \right) = 0$, $\sigma  = 1$ and solving Eq.\eqref{e3_3}
yield $\mu \left( x \right) =  - x - \ln \sqrt {e\left( {e - 1} \right)}  \approx  - x - 0.7707$ and $\delta \left( x \right) =  - \frac{{{e^{ - x}}}}{{\sqrt {e - 1} }} \approx  - 0.7629{e^{ - x}}$.

\begin{figure}[H]
	\begin{center}
		\includegraphics[width=0.8\textwidth]{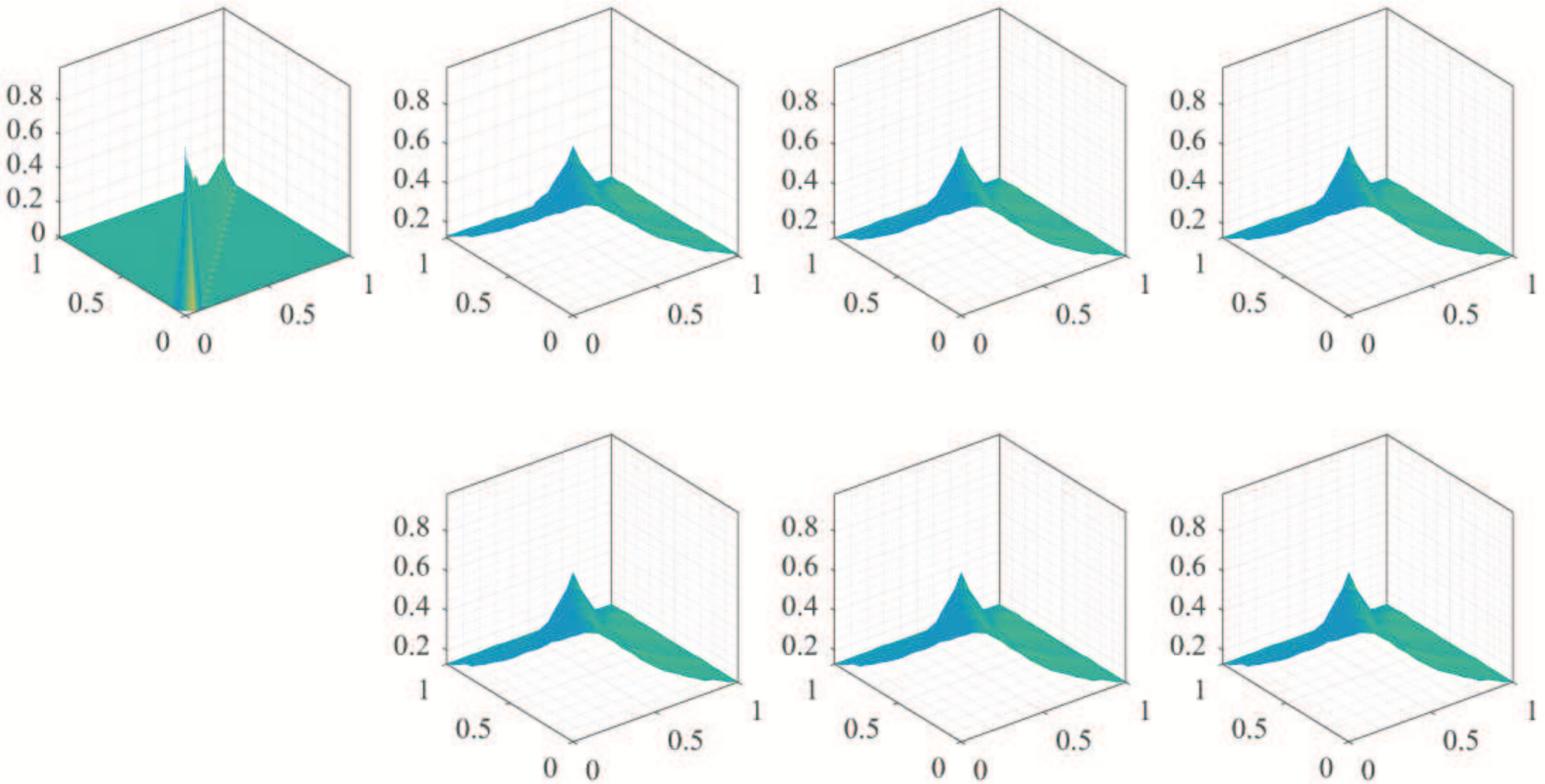}
		\caption{Iterations of sample covariance functions $\left\{ {{T^{\left( k \right)}}} \right\}_{k = 0}^6$.}
		\label{SP301}
	\end{center}
\end{figure}

\begin{figure}[H]
	\begin{center}
		\includegraphics[width=0.7\textwidth]{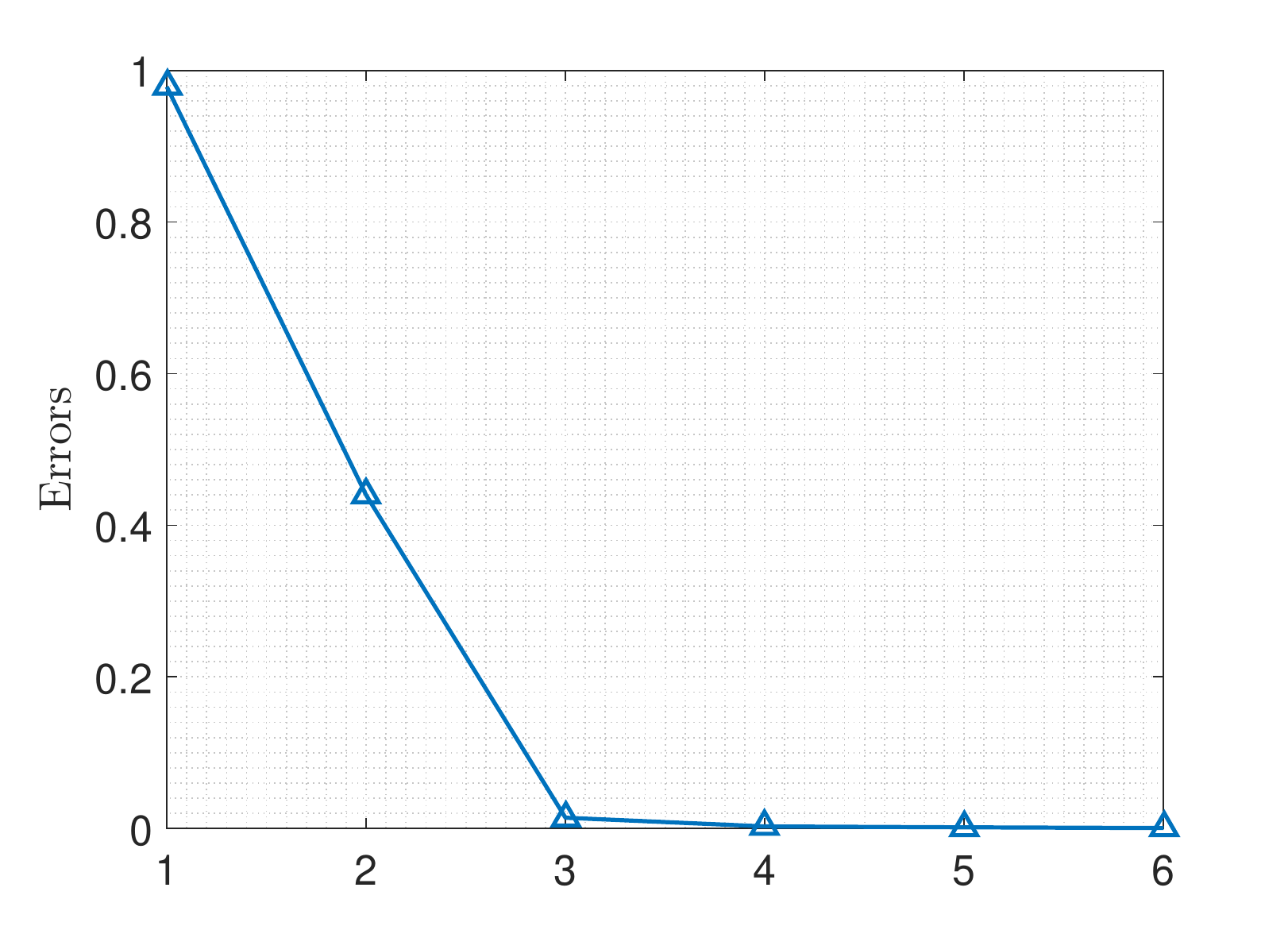}
		\caption{Iterative errors.}
		\label{SP302}
	\end{center}
\end{figure}

Fig.\ref{SP301} shows iterations of sample covariance functions $\left\{ {{T^{\left( k \right)}}} \right\}_{k = 0}^6$ and the corresponding convergence error of each iteration is shown as Fig.\ref{SP302}. The good convergence of Algorithm \ref{Alg: s1} for non-gaussian and non-stationary stochastic processes is demonstrated.

\begin{figure}[H]
	\begin{center}
		\includegraphics[width=1.0\textwidth]{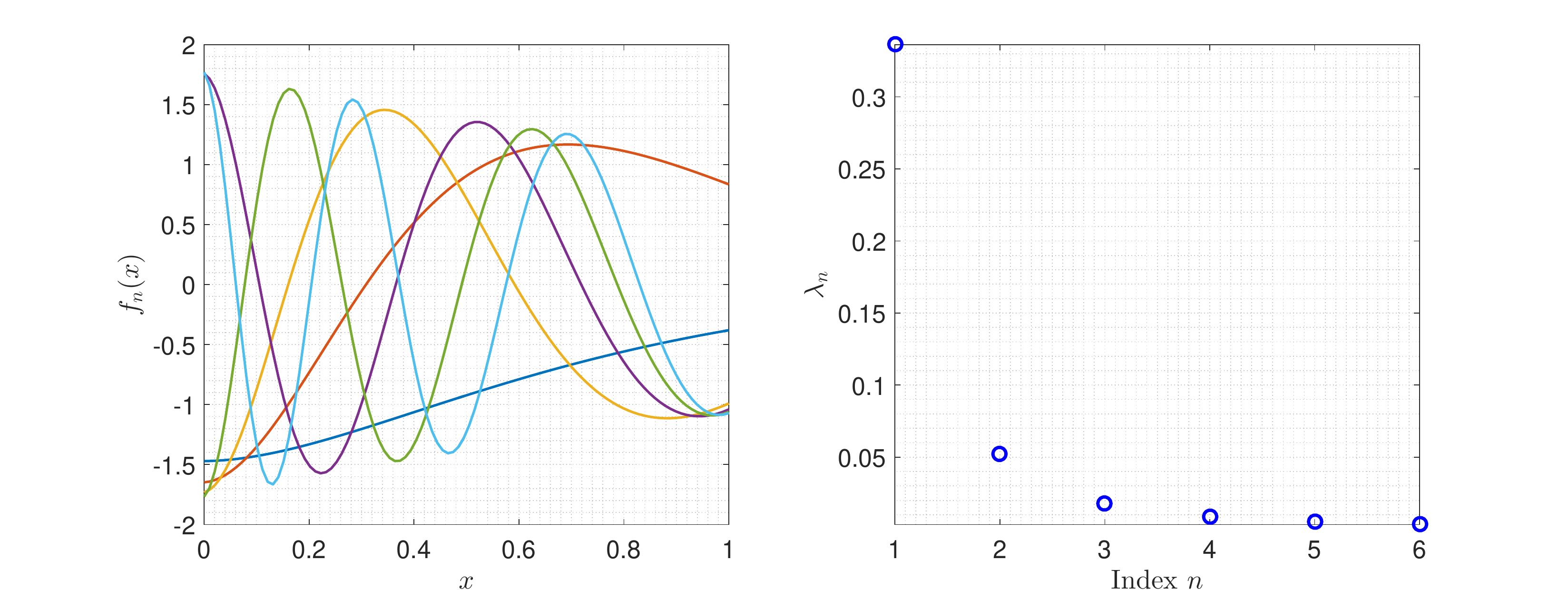}
		\caption{Eigenfunctions (left) and eigenvalues (right) of covariance $C\left( {{x_1},{x_2}} \right)$.}
		\label{SP303}
	\end{center}
\end{figure}

According to Eq.\eqref{s1_4}, first six eigenfunctions and eigenvalues of $C\left( {{x_1},{x_2}} \right)$ are shown in Fig.\ref{SP303}.

\begin{figure}[H]
	\begin{center}
		\includegraphics[width=0.7\textwidth]{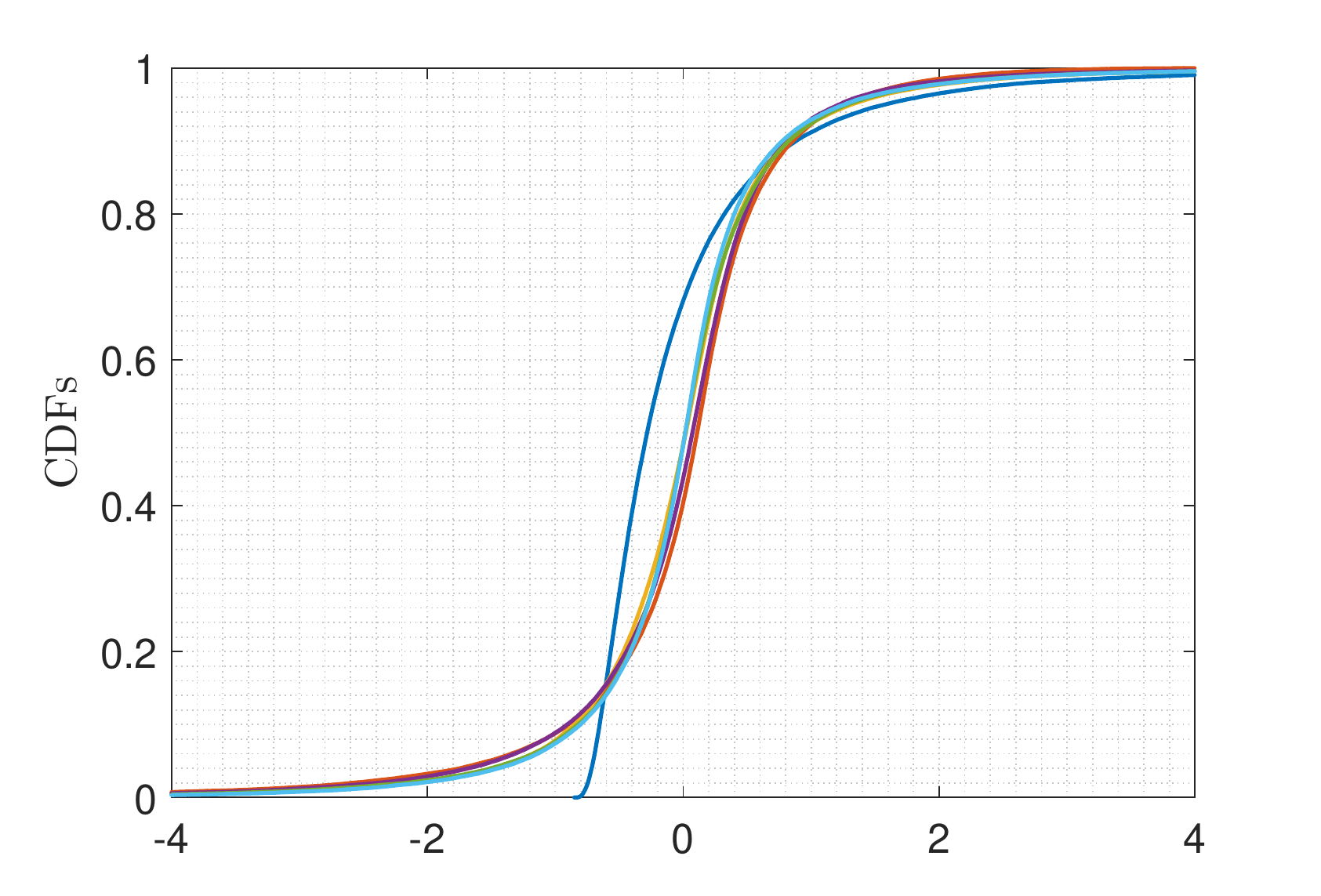}
		\caption{CDFs of random variables $\left\{ {{\xi _i}\left( \theta  \right)} \right\}_{i = 1}^6$.}
		\label{SP304}
	\end{center}
\end{figure}

\begin{figure}[H]
	\begin{center}
		\includegraphics[width=0.8\textwidth]{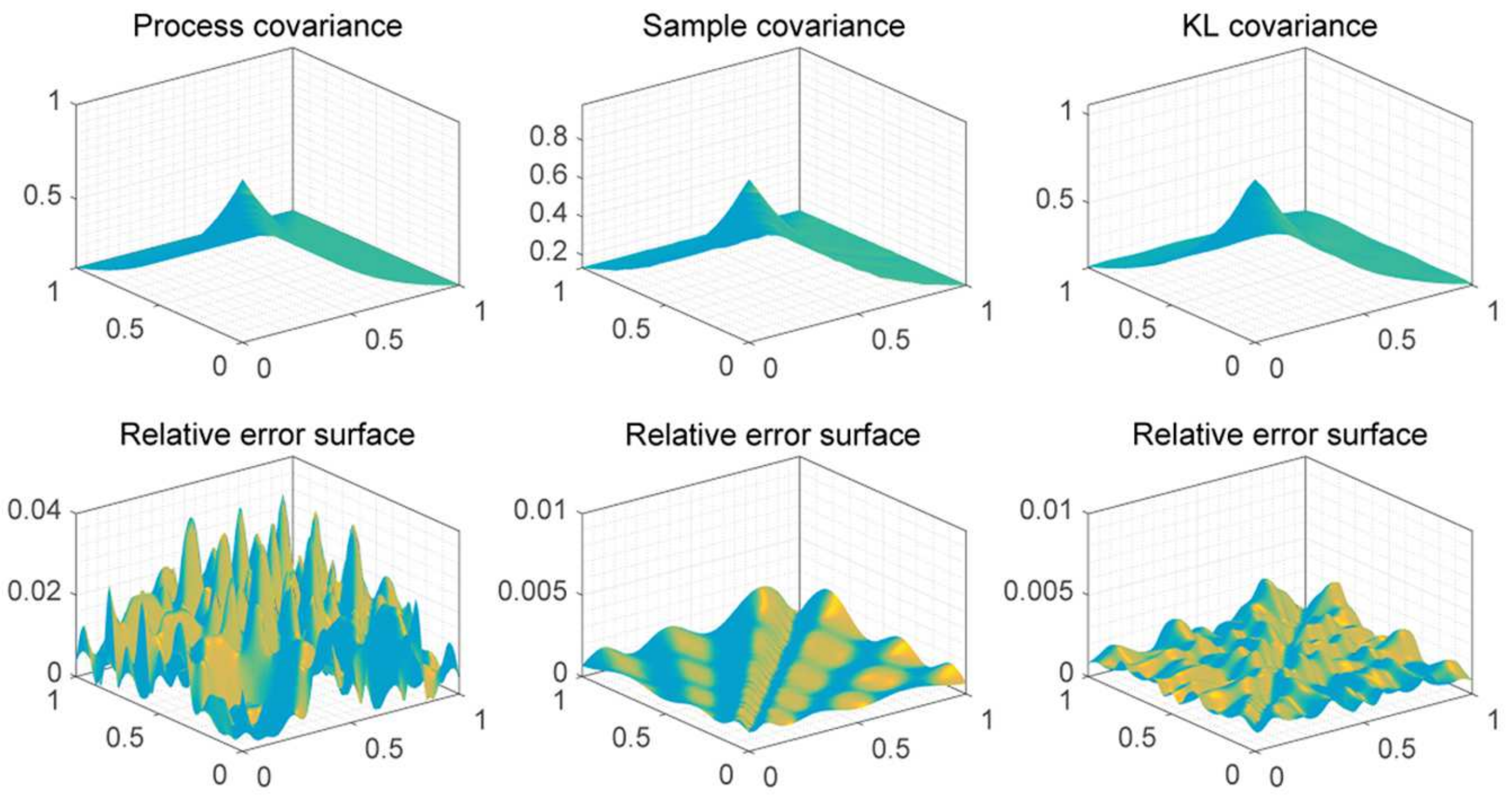}
		\caption{Comparisons between target, sample and KL-simulated covariance.}
		\label{SP305}
	\end{center}
\end{figure}

Fig.\ref{SP304} shows CDFs of random variables $\left\{ {{\xi _i}\left( \theta  \right)} \right\}_{i = 1}^6$ and their uncorrelated properties are shown as Table.\ref{E3t1}.
Fig.\ref{SP305} shows comparisons between target, sample and KL-simulated covariance (the corresponding legends of relative errors are the same as Fig.\ref{SP105}), which verify the applicability of the proposed Algorithm \ref{Alg: s1} to strongly non-gaussian and non-stationary stochastic processes.
\begin{table}[H]
     \centering
	\caption{Statistical correlations between ${\xi _i}\left( \theta  \right)$ and ${\xi _j}\left( \theta  \right)$}\vspace{-0.6em}
	\label{E3t1}
	\begin{tabular}{rrrrrrr}
		\toprule
		$E\left\{ {{\xi _i}{\xi _j}} \right\}$ & ${\xi _1}\left( \theta  \right)$ & ${\xi _2}\left( \theta  \right)$ & ${\xi _3}\left( \theta  \right)$ & ${\xi _4}\left( \theta  \right)$ & ${\xi _5}\left( \theta  \right)$ & ${\xi _6}\left( \theta  \right)$ \\
		\midrule
		${\xi _1}\left( \theta  \right)$ &  1.0051 \\
		${\xi _2}\left( \theta  \right)$ &  $-$0.0015  &  0.9979 \\
		${\xi _3}\left( \theta  \right)$ &  $-$0.0148  &  0.0014  & 1.0049 &     & sym. \\
		${\xi _4}\left( \theta  \right)$ &  0.0141  &  0.0004  & $-$0.0028 &  0.9988 \\
		${\xi _5}\left( \theta  \right)$ &  $-$0.0050  &  0.0066  & 0.0077 &  0.0038  &  0.9997	 \\
		${\xi _6}\left( \theta  \right)$ &  $-$0.0093   &  0.0069  & 0.0029 &   $-$0.0076 &  0.0061 &  1.0042 \\
		\bottomrule
	\end{tabular}
\end{table}

\section{Conclusion}
In this paper, efficient numerical schemes have been presented for simulating non-gaussian and non-stationary stochastic processes specified by covariance functions and marginal distribution functions. In order to simulate samples of the target stochastic process, stochastic samples automatically matching the target marginal distribution function are firstly generated, and an iterative algorithm is proposed to match the target covariance function by transform the order of initial stochastic samples. Three numerical examples demonstrate the fast convergence and the high accuracy of the proposed algorithm. In order to overcome the difficulty that sample-descriptions are not convenient to applied to subsequent stochastic analysis, three numerical algorithms are developed to represent the obtained stochastic samples based on KL expansion and PC expansion. Different algorithms can be used for different problems of practical interests and the performances of the developed algorithms are indicated by numerical examples. All proposed algorithms can be readily extended to multi-dimensional random fields and will be shown in subsequent researches.

\section*{Acknowledgments}
This research was supported by the National Natural Science Foundation of China (Project 11972009). This support is gratefully acknowledged.

\bibliography{References}

\begin{thebibliography}{32}
\expandafter\ifx\csname natexlab\endcsname\relax\def\natexlab#1{#1}\fi
\providecommand{\url}[1]{\texttt{#1}}
\providecommand{\href}[2]{#2}
\providecommand{\path}[1]{#1}
\providecommand{\DOIprefix}{doi:}
\providecommand{\ArXivprefix}{arXiv:}
\providecommand{\URLprefix}{URL: }
\providecommand{\Pubmedprefix}{pmid:}
\providecommand{\doi}[1]{\href{http://dx.doi.org/#1}{\path{#1}}}
\providecommand{\Pubmed}[1]{\href{pmid:#1}{\path{#1}}}
\providecommand{\bibinfo}[2]{#2}
\ifx\xfnm\relax \def\xfnm[#1]{\unskip,\space#1}\fi
\bibitem[{Ghanem and Spanos(2003)}]{ghanem2003stochastic}
\bibinfo{author}{R.~G. Ghanem}, \bibinfo{author}{P.~D. Spanos},
  \bibinfo{title}{Stochastic finite elements: a spectral approach},
  \bibinfo{publisher}{Courier Corporation}, \bibinfo{year}{2003}.
\bibitem[{Tankov(2003)}]{tankov2003financial}
\bibinfo{author}{P.~Tankov}, \bibinfo{title}{Financial modelling with jump
  processes}, \bibinfo{publisher}{Chapman and Hall/CRC}, \bibinfo{year}{2003}.
\bibitem[{Aboy et~al.(2005)Aboy, M{\'a}rquez, McNames, Hornero, Trong, and
  Goldstein}]{aboy2005adaptive}
\bibinfo{author}{M.~Aboy}, \bibinfo{author}{O.~W. M{\'a}rquez},
  \bibinfo{author}{J.~McNames}, \bibinfo{author}{R.~Hornero},
  \bibinfo{author}{T.~Trong}, \bibinfo{author}{B.~Goldstein},
\newblock \bibinfo{title}{Adaptive modeling and spectral estimation of
  nonstationary biomedical signals based on kalman filtering},
\newblock \bibinfo{journal}{IEEE Transactions on Biomedical Engineering}
  \bibinfo{volume}{52} (\bibinfo{year}{2005}) \bibinfo{pages}{1485--1489}.
\bibitem[{Stefanou(2009)}]{stefanou2009stochastic}
\bibinfo{author}{G.~Stefanou},
\newblock \bibinfo{title}{The stochastic finite element method: past, present
  and future},
\newblock \bibinfo{journal}{Computer methods in applied mechanics and
  engineering} \bibinfo{volume}{198} (\bibinfo{year}{2009})
  \bibinfo{pages}{1031--1051}.
\bibitem[{Papoulis and Pillai(2002)}]{papoulis2002probability}
\bibinfo{author}{A.~Papoulis}, \bibinfo{author}{S.~U. Pillai},
  \bibinfo{title}{Probability, random variables, and stochastic processes},
  \bibinfo{publisher}{Tata McGraw-Hill Education}, \bibinfo{year}{2002}.
\bibitem[{Grigoriu(2006)}]{grigoriu2006evaluation}
\bibinfo{author}{M.~Grigoriu},
\newblock \bibinfo{title}{Evaluation of karhunen--lo{\`e}ve, spectral, and
  sampling representations for stochastic processes},
\newblock \bibinfo{journal}{Journal of engineering mechanics}
  \bibinfo{volume}{132} (\bibinfo{year}{2006}) \bibinfo{pages}{179--189}.
\bibitem[{Rasmussen(2003)}]{rasmussen2003gaussian}
\bibinfo{author}{C.~E. Rasmussen},
\newblock \bibinfo{title}{Gaussian processes in machine learning},
\newblock in: \bibinfo{booktitle}{Summer School on Machine Learning},
  \bibinfo{organization}{Springer}, \bibinfo{year}{2003}, pp.
  \bibinfo{pages}{63--71}.
\bibitem[{Phoon et~al.(2004)Phoon, Huang, and Quek}]{phoon2004comparison}
\bibinfo{author}{K.~Phoon}, \bibinfo{author}{H.~Huang},
  \bibinfo{author}{S.~Quek},
\newblock \bibinfo{title}{Comparison between karhunen--loeve and wavelet
  expansions for simulation of gaussian processes},
\newblock \bibinfo{journal}{Computers \& structures} \bibinfo{volume}{82}
  (\bibinfo{year}{2004}) \bibinfo{pages}{985--991}.
\bibitem[{Deodatis and Micaletti(2001)}]{deodatis2001simulation}
\bibinfo{author}{G.~Deodatis}, \bibinfo{author}{R.~C. Micaletti},
\newblock \bibinfo{title}{Simulation of highly skewed non-gaussian stochastic
  processes},
\newblock \bibinfo{journal}{Journal of engineering mechanics}
  \bibinfo{volume}{127} (\bibinfo{year}{2001}) \bibinfo{pages}{1284--1295}.
\bibitem[{Yamazaki and Shinozuka(1988)}]{yamazaki1988digital}
\bibinfo{author}{F.~Yamazaki}, \bibinfo{author}{M.~Shinozuka},
\newblock \bibinfo{title}{Digital generation of {non-Gaussian} stochastic
  fields},
\newblock \bibinfo{journal}{Journal of Engineering Mechanics}
  \bibinfo{volume}{114} (\bibinfo{year}{1988}) \bibinfo{pages}{1183--1197}.
\bibitem[{Shinozuka and Deodatis(1991)}]{shinozuka1991simulation}
\bibinfo{author}{M.~Shinozuka}, \bibinfo{author}{G.~Deodatis},
\newblock \bibinfo{title}{Simulation of stochastic processes by spectral
  representation},
\newblock \bibinfo{journal}{Applied Mechanics Reviews} \bibinfo{volume}{44}
  (\bibinfo{year}{1991}) \bibinfo{pages}{191--204}.
\bibitem[{Elishakoff et~al.(1994)Elishakoff, Ren, and
  Shinozuka}]{Elishakoff1994Conditional}
\bibinfo{author}{I.~Elishakoff}, \bibinfo{author}{Y.~J. Ren},
  \bibinfo{author}{M.~Shinozuka},
\newblock \bibinfo{title}{Conditional simulation of non-gaussian random
  fields},
\newblock \bibinfo{journal}{Engineering Structures} \bibinfo{volume}{16}
  (\bibinfo{year}{1994}) \bibinfo{pages}{558--563}.
\bibitem[{Popescu et~al.(1998)Popescu, Deodatis, and
  Prevost}]{Popescu1998Simulation}
\bibinfo{author}{R.~Popescu}, \bibinfo{author}{G.~Deodatis},
  \bibinfo{author}{J.~H. Prevost},
\newblock \bibinfo{title}{Simulation of homogeneous nongaussian stochastic
  vector fields},
\newblock \bibinfo{journal}{Probabilistic Engineering Mechanics}
  \bibinfo{volume}{13} (\bibinfo{year}{1998}) \bibinfo{pages}{1--13}.
\bibitem[{Grigoriu(1998)}]{grigoriu1998simulation}
\bibinfo{author}{M.~Grigoriu},
\newblock \bibinfo{title}{Simulation of stationary {non-Gaussian} translation
  processes},
\newblock \bibinfo{journal}{Journal of engineering mechanics}
  \bibinfo{volume}{124} (\bibinfo{year}{1998}) \bibinfo{pages}{121--126}.
\bibitem[{Liu et~al.(2016)Liu, Liu, and Peng}]{liu2016random}
\bibinfo{author}{Z.~Liu}, \bibinfo{author}{W.~Liu}, \bibinfo{author}{Y.~Peng},
\newblock \bibinfo{title}{Random function based spectral representation of
  stationary and non-stationary stochastic processes},
\newblock \bibinfo{journal}{Probabilistic Engineering Mechanics}
  \bibinfo{volume}{45} (\bibinfo{year}{2016}) \bibinfo{pages}{115--126}.
\bibitem[{Huang et~al.(2001)Huang, Quek, and Phoon}]{huang2001convergence}
\bibinfo{author}{S.~Huang}, \bibinfo{author}{S.~Quek},
  \bibinfo{author}{K.~Phoon},
\newblock \bibinfo{title}{Convergence study of the truncated karhunen--loeve
  expansion for simulation of stochastic processes},
\newblock \bibinfo{journal}{International journal for numerical methods in
  engineering} \bibinfo{volume}{52} (\bibinfo{year}{2001})
  \bibinfo{pages}{1029--1043}.
\bibitem[{Sudret and Der~Kiureghian(2000)}]{sudret2000stochastic}
\bibinfo{author}{B.~Sudret}, \bibinfo{author}{A.~Der~Kiureghian},
\newblock \bibinfo{title}{Stochastic finite elements and reliability: a
  state-of-the-art report},
\newblock \bibinfo{journal}{University of California, Berkeley}
  (\bibinfo{year}{2000}) \bibinfo{pages}{114--120}.
\bibitem[{Poirion and Zentner(2013)}]{poirion2013non}
\bibinfo{author}{F.~Poirion}, \bibinfo{author}{I.~Zentner},
\newblock \bibinfo{title}{Non-gaussian non-stationary models for natural hazard
  modeling},
\newblock \bibinfo{journal}{Applied Mathematical Modelling}
  \bibinfo{volume}{37} (\bibinfo{year}{2013}) \bibinfo{pages}{5938--5950}.
\bibitem[{Kim and Shields(2015)}]{kim2015modeling}
\bibinfo{author}{H.~Kim}, \bibinfo{author}{M.~D. Shields},
\newblock \bibinfo{title}{Modeling strongly non-gaussian non-stationary
  stochastic processes using the {Iterative Translation Approximation Method}
  and {Karhunen--Lo{\`e}ve} expansion},
\newblock \bibinfo{journal}{Computers \& Structures} \bibinfo{volume}{161}
  (\bibinfo{year}{2015}) \bibinfo{pages}{31--42}.
\bibitem[{Liu et~al.(2017)Liu, Liu, and Peng}]{liu2017dimension}
\bibinfo{author}{Z.~Liu}, \bibinfo{author}{Z.~Liu}, \bibinfo{author}{Y.~Peng},
\newblock \bibinfo{title}{Dimension reduction of {Karhunen-Loeve} expansion for
  simulation of stochastic processes},
\newblock \bibinfo{journal}{Journal of Sound and Vibration}
  \bibinfo{volume}{408} (\bibinfo{year}{2017}) \bibinfo{pages}{168--189}.
\bibitem[{Phoon et~al.(2002)Phoon, Huang, and Quek}]{phoon2002simulation}
\bibinfo{author}{K.~Phoon}, \bibinfo{author}{S.~Huang},
  \bibinfo{author}{S.~Quek},
\newblock \bibinfo{title}{Simulation of second-order processes using
  {Karhunen--Loeve} expansion},
\newblock \bibinfo{journal}{Computers \& structures} \bibinfo{volume}{80}
  (\bibinfo{year}{2002}) \bibinfo{pages}{1049--1060}.
\bibitem[{Phoon et~al.(2005)Phoon, Huang, and Quek}]{phoon2005simulation}
\bibinfo{author}{K.~Phoon}, \bibinfo{author}{H.~Huang},
  \bibinfo{author}{S.~Quek},
\newblock \bibinfo{title}{Simulation of strongly {non-Gaussian} processes using
  {Karhunen--Loeve} expansion},
\newblock \bibinfo{journal}{Probabilistic Engineering Mechanics}
  \bibinfo{volume}{20} (\bibinfo{year}{2005}) \bibinfo{pages}{188--198}.
\bibitem[{Dai et~al.(2019)Dai, Zheng, and Ma}]{dai2019explicit}
\bibinfo{author}{H.~Dai}, \bibinfo{author}{Z.~Zheng}, \bibinfo{author}{H.~Ma},
\newblock \bibinfo{title}{An explicit method for simulating non-gaussian and
  non-stationary stochastic processes by {Karhunen--Lo{\`e}ve} and polynomial
  chaos expansion},
\newblock \bibinfo{journal}{Mechanical Systems and Signal Processing}
  \bibinfo{volume}{115} (\bibinfo{year}{2019}) \bibinfo{pages}{1--13}.
\bibitem[{Sakamoto and Ghanem(2002{\natexlab{a}})}]{sakamoto2002polynomial}
\bibinfo{author}{S.~Sakamoto}, \bibinfo{author}{R.~Ghanem},
\newblock \bibinfo{title}{Polynomial chaos decomposition for the simulation of
  {non-Gaussian} nonstationary stochastic processes},
\newblock \bibinfo{journal}{Journal of engineering mechanics}
  \bibinfo{volume}{128} (\bibinfo{year}{2002}{\natexlab{a}})
  \bibinfo{pages}{190--201}.
\bibitem[{Sakamoto and Ghanem(2002{\natexlab{b}})}]{sakamoto2002simulation}
\bibinfo{author}{S.~Sakamoto}, \bibinfo{author}{R.~Ghanem},
\newblock \bibinfo{title}{Simulation of multi-dimensional non-gaussian
  non-stationary random fields},
\newblock \bibinfo{journal}{Probabilistic Engineering Mechanics}
  \bibinfo{volume}{17} (\bibinfo{year}{2002}{\natexlab{b}})
  \bibinfo{pages}{167--176}.
\bibitem[{Puig et~al.(2002)Puig, Poirion, and Soize}]{puig2002non}
\bibinfo{author}{B.~Puig}, \bibinfo{author}{F.~Poirion},
  \bibinfo{author}{C.~Soize},
\newblock \bibinfo{title}{{Non-Gaussian} simulation using hermite polynomial
  expansion: convergences and algorithms},
\newblock \bibinfo{journal}{Probabilistic Engineering Mechanics}
  \bibinfo{volume}{17} (\bibinfo{year}{2002}) \bibinfo{pages}{253--264}.
\bibitem[{Field~Jr and Grigoriu(2004)}]{field2004accuracy}
\bibinfo{author}{R.~Field~Jr}, \bibinfo{author}{M.~Grigoriu},
\newblock \bibinfo{title}{On the accuracy of the polynomial chaos
  approximation},
\newblock \bibinfo{journal}{Probabilistic Engineering Mechanics}
  \bibinfo{volume}{19} (\bibinfo{year}{2004}) \bibinfo{pages}{65--80}.
\bibitem[{Lagaros et~al.(2005)Lagaros, Stefanou, and
  Papadrakakis}]{lagaros2005enhanced}
\bibinfo{author}{N.~D. Lagaros}, \bibinfo{author}{G.~Stefanou},
  \bibinfo{author}{M.~Papadrakakis},
\newblock \bibinfo{title}{An enhanced hybrid method for the simulation of
  highly skewed non-gaussian stochastic fields},
\newblock \bibinfo{journal}{Computer Methods in Applied Mechanics and
  Engineering} \bibinfo{volume}{194} (\bibinfo{year}{2005})
  \bibinfo{pages}{4824--4844}.
\bibitem[{Zheng and Dai(2017)}]{zheng2017simulation}
\bibinfo{author}{Z.~Zheng}, \bibinfo{author}{H.~Dai},
\newblock \bibinfo{title}{Simulation of multi-dimensional random fields by
  {Karhunen--Lo{\`e}ve} expansion},
\newblock \bibinfo{journal}{Computer Methods in Applied Mechanics and
  Engineering} \bibinfo{volume}{324} (\bibinfo{year}{2017})
  \bibinfo{pages}{221--247}.
\bibitem[{Betz et~al.(2014)Betz, Papaioannou, and Straub}]{betz2014numerical}
\bibinfo{author}{W.~Betz}, \bibinfo{author}{I.~Papaioannou},
  \bibinfo{author}{D.~Straub},
\newblock \bibinfo{title}{Numerical methods for the discretization of random
  fields by means of the {Karhunen--Lo{\`e}ve} expansion},
\newblock \bibinfo{journal}{Computer Methods in Applied Mechanics and
  Engineering} \bibinfo{volume}{271} (\bibinfo{year}{2014})
  \bibinfo{pages}{109--129}.
\bibitem[{Christou et~al.(2016)Christou, Bocchini, and
  Miranda}]{christou2016optimal}
\bibinfo{author}{V.~Christou}, \bibinfo{author}{P.~Bocchini},
  \bibinfo{author}{M.~J. Miranda},
\newblock \bibinfo{title}{Optimal representation of multi-dimensional random
  fields with a moderate number of samples: {Application} to stochastic
  mechanics},
\newblock \bibinfo{journal}{Probabilistic Engineering Mechanics}
  \bibinfo{volume}{44} (\bibinfo{year}{2016}) \bibinfo{pages}{53--65}.
\bibitem[{Xiu(2010)}]{xiu2010numerical}
\bibinfo{author}{D.~Xiu}, \bibinfo{title}{Numerical methods for stochastic
  computations: a spectral method approach}, \bibinfo{publisher}{Princeton
  university press}, \bibinfo{year}{2010}.

\end{thebibliography}

\end{document}